\documentclass[sigplan,10pt,nonacm]{acmart}
\usepackage{ulem}
\usepackage{pifont}
\usepackage{graphicx}
\usepackage{subfloat}
\usepackage{tabularray}
\usepackage{bbding}
\usepackage{makecell}
\usepackage{cleveref}
\usepackage{amsmath}
\usepackage{tikz}
\usepackage{multirow}
\usepackage{subcaption}
\usepackage{xcolor}
\usepackage[linesnumbered, ruled] 
{algorithm2e}
\usepackage{pifont}
\newcommand*\circled[1]{\tikz[baseline=(char.base)]{
            \node[circle,fill=.,inner sep=0.8pt] (char) {\textcolor{white}{#1}};}}

\newcommand*\whitecircled[1]{\tikz[baseline=(char.base)]{
            \node[shape=circle, draw, inner sep=1pt] (char) {#1};}}

\newcommand\code[1]{\texttt{#1}}

\newcommand*{\Mname}{PatchedServe}

\begin{document}

\title{\Mname: A Patch Management Framework for SLO-Optimized Hybrid Resolution Diffusion Serving}

\author{Desen Sun}
\affiliation{%
  \institution{University of Waterloo}
  \city{Waterloo}
  \state{Ontario}
  \country{Canada}
}
% \email{your_email@uwaterloo.ca}

\author{Zepeng Zhao}
\affiliation{%
  \institution{Carnegie Mellon University}
  \city{Pittsburgh}
  \state{Pennsylvania}
  \country{US}
}
% \email{}

\author{Yuke Wang}
\affiliation{%
  \institution{Rice University}
  \city{Houston}
  \state{Texas}
  \country{US}
}
% \email{}

% \author{{\rm Desen Sun, Zepeng Zhao, and Yuke Wang}\\University of Waterloo}
% \author[1]{Desen Sun} 
% \author[2]{Zepeng Zhao} 
% \author[3]{Yuke Wang} 
% \affil[1]{University of Waterloo}
% \affil[2]{Carnegie Mellon University}
% \affil[3]{Rice University}

\fancyhead{}  % clears the header fields that trigger HotCRP format checker
\renewcommand\footnotetextcopyrightpermission[1]{} % removes the ACM copyright footnote

\begin{abstract}
The Text-to-Image (T2I) diffusion model has emerged as one of the most widely adopted generative models. However, serving diffusion models at the granularity of entire images introduces significant challenges, particularly under multi-resolution workloads. First, image-level serving obstructs batching across requests. Second, heterogeneous resolutions exhibit distinct locality characteristics, making it difficult to apply a uniform cache policy effectively.

To address these challenges, we present \Mname, a Patch Management Framework for SLO-Optimized Hybrid-Resolution Diffusion Serving. \Mname{} introduces a patch-level management strategy that enables batching across heterogeneous resolutions. Specifically, it incorporates a novel patch-based processing workflow that substantially improves throughput for hybrid-resolution inputs. Moreover, \Mname{} devises a patch-level cache reuse policy to fully exploit diffusion redundancies, and integrates an SLO-aware scheduling algorithm with lightweight online latency prediction to improve responsiveness.
Our evaluation demonstrates that \Mname{} achieves 30.1\,\% higher SLO satisfaction than the state-of-the-art diffusion serving system, while preserving image quality.
\end{abstract}

\maketitle
\pagestyle{plain}

\section{Introduction}
Over recent years, Text-to-Image (T2I) diffusion models have become increasingly popular \cite{NEURIPS2021_49ad23d1} and have been instrumental for many companies, including Google Imagen  \cite{NEURIPS2022_ec795aea}, Adobe Firefly \cite{firely}, OpenAI DALLE \cite{dalle3}, etc. Users have exploited diffusion models for various purposes, such as designing scenes \cite{Bokhovkin_2025_CVPR,Huang_2025_CVPR_MIDI,Yang_2025_CVPR_Prometheus,eldesokey2025buildascene}, characters \cite{Sun_2025_CVPR_DRiVE,Wang_ACM_MM_2024_Evolving}, or posters \cite{Chen_2025_CVPR_POSTA,Gao_2025_CVPR_PosterMaker,Ma_Deng_Chen_Du_Lu_Yang_2025}, primarily due to their ability to generate images with superior quality. T2I diffusion models synthesize images from Gaussian noise by iteratively denoising, adopting both ResNet \cite{he2016deep} and Transformer \cite{vaswani2017attention} architectures.

Unlike LLMs, which can efficiently handle variable sequence lengths through KV caching \cite{vllm}, diffusion models require multiple steps of computation-intensive attention without reusable KV caches. Consequently, heterogeneity in tensor shapes propagates throughout the entire diffusion pipeline, limiting parallel execution opportunities and preventing the system from scaling batch size. For example, generating three SDXL requests of different resolutions ($512\times512$, $768\times768$, and $1024\times1024$) on H100 executes in 9.5s when processing them concurrently in a batch, compared to 17.8s when executed sequentially. In real-world serving scenarios, users frequently request images of different resolutions for diverse application needs \cite{podell2023sdxl, dalle3, hunyuan, mdm, ControlAR}, leading to long waiting times as disability to handle mixed resolution requests simultaneously.

Prior studies have explored optimizing the performance of diffusion models.
Some works \cite{li2023distrifusion,fang2024pipefusionpatchlevelpipelineparallelism,xdit} exploit patch parallelism to reduce latency on multiple GPUs. Another category of studies seek to reuse the cache to accelerate inference \cite{blockcache,nirvana,deepcache,qiu2025acceleratingdiffusiontransformergradientoptimized,yu2025abcachetrainingfreeaccelerationdiffusion,EXION,cheng2025catpruningclusterawaretoken}. While these methods effectively reduce request latency and mitigate waiting times, they remain insufficient to achieve a high Service Level Objective (SLO) which is necessary in building a strong serving system, as they overlook the complexities introduced by mixed-resolution configurations.

\begin{figure}[t]

  \begin{center}
  \includegraphics[width=1\linewidth]{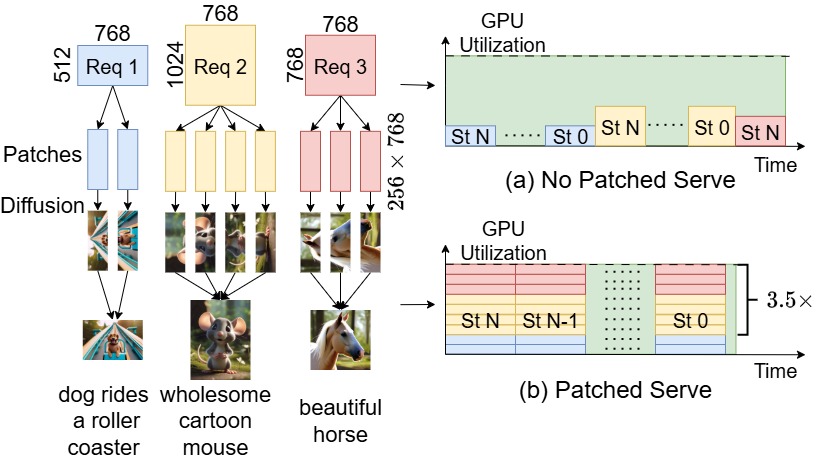}
\caption{\label{fig:intro} Assume three requests, Req1, Req2, and Req3, where each requiring processing over N steps, from St N to St 0. (a) Process requests sequentially. (b) Process requests in parallel, achieving higher GPU utilization.}
  \end{center}
\end{figure}

With the above insights, we propose a patch-level parallelization strategy that restructures heterogeneous diffusion workloads into uniform computational units.
Our key observation is that most operations in diffusion models (e.g., \code{Linear}, \code{FeedForward}, and \code{Cross Attention}) primarily operate on ``local'' (pixel level) rather than ``global'' (image level) information. Although originally designed to process full images, these operations can be decomposed into smaller sub-operations over individual patches. Once patches share identical shapes, heterogeneous requests can be combined as a single batch, converting resolution diversity into parallelizable work.
For a simple illustration ({\Cref{fig:intro}), 
there are 3 requests with different resolutions. Without customization, these requests must be processed sequentially due to mismatched input shapes (\Cref{fig:intro}(a)), leading to underutilized GPUs. In contrast, by segmenting the requests into fine-grained patches with uniform shapes, they can be processed concurrently in a single batch (\Cref{fig:intro}(b)), significantly improving parallel efficiency.
% In this way, the system can reach 3.5$\times$ higher GPU utilization.
\iffalse
With patching, the diffusion serving can exploit abundant locality and parallelism benefits across different levels of compute and memory hierarchies, including inter-patch computational parallelism and intra-patch data locality, to reducing the overall latency.
\fi
% Obviously, an image of size 1024 $\times$ 1024 can be divided into four 512 $\times$ 512 sub-images. Based on this division, we can change such 1024 $\times$ 1024 image synthesis task to four 512 $\times$ 512 ones, which can be processed together with another 512 $\times$ 512 request instead of computing two requests sequentially. We call such sub-image as patch and patch size denotes the dimensions of each sub-image. By dividing the image, ESyMReD can form a 5 images batch as the  model input, effectively utilizing the wasted GPU resource and improving system throughput. After completing all diffusion steps, ESyMReD generates five 512 $\times$ 512 images: one for a complete image and 4 sub-images from different parts of a large image. ESyMReD further merges these four sub-images into a single image as the result of Request 2.
}

% \todo{Different from those prior efforts, we spot another new direction of addressing the issue by ....Our key observation is ....}
% \todo{I think here is missing one paragraph to illustrate the potential observations that why dividing the image as patch could help with our design to address the GPU underutilization problem.}

% \todo{\textbf{I strongly suggest put these challenges and the Figure-1 ahead of ``To this end, we propose....'' This paragraph should only talk about the key highlight of our proposed solution and how it addresses the each of these issues effectively.}}

{Despite patch-level decomposition enables higher degrees of parallelism, several challenges still prevent us from fully achieving these benefits.}
{\textbf{First}, partitioning an image introduces cross-patch dependencies. For example, \texttt{Convolution} operator in U-Net based diffusion models aggregates information from adjacent pixels at each location. If we split images naively, computations near patch boundaries become inaccurate because each patch lacks the adjacent pixels that would normally be included if the entire image were processed as a whole. 
\iffalse
For example, \texttt{GroupNorm} operations in U-Net-based diffusion models will calculate the mean and variance values of the entire image. Although it is easy to calculate these metrics for individual patches, it is difficult to calculate them for the whole image due to the extensive exchange of partial results across patches.} 
\fi
% Notice that exchanging such information independently will incur non-negligible overhead. 
% Therefore it is necessary to design a fusion kernel that mitigates such overhead by overlapping it with other operations, leveraging the parallelism capability of GPU.
% \item 
{\textbf{Second}, patch-level locality leveraging can introduce severe overhead. Cache-based mechanisms can further enhance the performance of diffusion models. However, combining patch-level processing with caching is challenging due to the additional overhead introduced by fine-grained cache management online.} Specifically, the reuse decisions must be made for each patch in every iteration, resulting in hundreds of decisions for a single image generation. And cache management operations (e.g.,  \code{insert}, \code{delete}, \code{query}) also require careful design.
\iffalse
To effectively reuse the patch-level cache while preserving image quality, the system must identify which patch outputs are sufficiently similar to cached results from the previous computation. These decisions must be made online before processing each block, and this process is repeated hundreds of times during generation. Even worse, cache management operations (such as \code{insert}, \code{delete}, \code{query}, and \code{re-organize}) are not light weight, demanding sophisticated design and optimization.}
\fi  
\textbf{Third}, diverse SLO requirements further complicate scheduling. In practice, service vendors schedule tasks by their deadline to maximize SLO satisfaction. However, the scheduling algorithm is hard to design due to challenging latency prediction, which depends on both batch size and resolution combinations, leading to substantial search space.

% \iffalse
% For instance, if a task with a higher resolution arrives earlier than one with a lower resolution, it is permissible for the larger task to be started later. The more candidate resolutions there are, the more diverse the SLO requirements will be. Such diversity will enlarge the scheduling searching space significantly, makes us hard to find the optimal solution.
% \fi

\begin{figure}[t]
  \begin{center}
  \includegraphics[width=\linewidth]{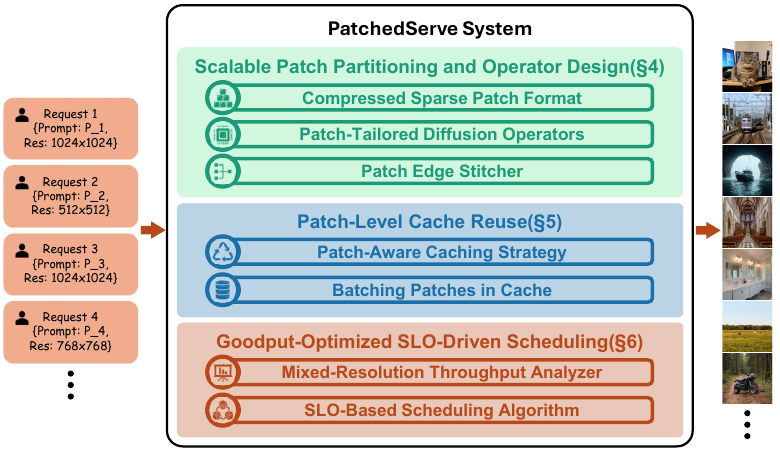}
  \end{center}
  \caption{\label{fig:overview} Overview of \Mname.}
\end{figure}

% \yuke{rewrite the entire paragraph!!!} 
To address these problems, we propose \Mname, a serving system that exploits fine-grained patch-level parallelism to enable efficient batching of mixed-resolution requests (\Cref{fig:overview}). {The design of \Mname{} comprises several key components: (1) A Patched inference mechanism (\S\ref{sec:patch-batch}). To enhance batch size, we first identify operations that require cross-patch context and then introduce two mechanisms that enable patched inference without hurting quality. \Mname~utilizes a novel patch management format to efficiently determine patch positions. Additionally, it incorporates a boundary stitcher to overlap memory movement overhead. (2) A patch level cache manager (\S\ref{sec:patch-cache}). To fully exploit the locality inherent in diffusion models, \Mname~employs a cache manager that determines reuse selection at the patch granularity. The manager coalesces fine-grained cache operations into batches for simultaneous execution, thereby improving parallel throughput. (3) An SLO-aware scheduling algorithm (\S\ref{sec:schedule}). \Mname~integrates an SLO-aware scheduler that maximizes SLO satisfaction under heterogeneous workloads. To support optimal decisions based on task latency, we further introduce a precise latency predictor. }

To sum up, we make the following contributions:
\begin{itemize}

% \item We propose a novel patch-based batching technique which enables requests with diverse resolutions to be processed together. To bring more improvement without losing the context information, we also implement a fused kernel to alleviate the extra overhead brought by exchanging information among patches.
\item {We propose a novel patch-based decomposition and batching strategy for mixed-resolution diffusion workloads. This approach not only enhances parallelism  but also preserves critical context information.}

\item {We introduce a patch-specific online cache management policy tailored to fully exploit the abundant patch-level locality efficiently.} 

\item We design a patch-aware scheduling algorithm that coalesces patch tasks with online latency prediction, achieving superior SLO satisfaction and goodput.

\item {Evaluation demonstrates that \Mname~improves parallel efficiency and achieves higher SLO satisfaction compared to existing approaches. As far as we know, \Mname{} is the first SLO-optimized T2I diffusion serving framework designed to handle hybrid-resolution requests.
% achieve up to 35.9\% tail latency reduction, up to 51\% higher SLO satisfaction rates, and 4.9$\times$ higher goodput with minimum quality loss compared with the state-of-the-art solutions.
}
\end{itemize}

\section{Background}
\subsection{Diffusion Models}

\begin{figure}[t]
  \begin{center}
  \includegraphics[width=\linewidth]{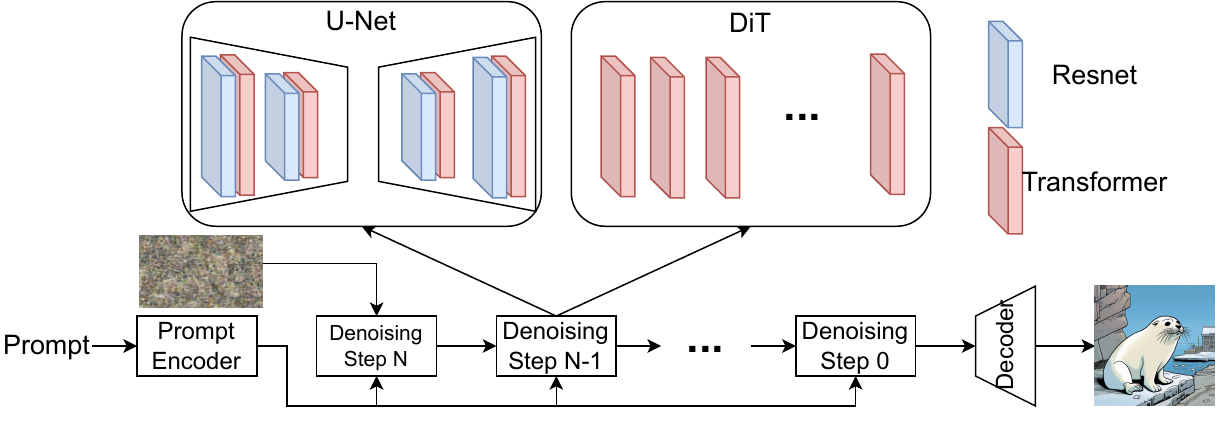}
  \end{center}
  \caption{\label{fig:diffusion} Latent Diffusion Model Structure. Two main types of backbones in the Diffusion model: U-Net and Diffusion Transformer (DiT).}
\end{figure}

Text-to-Image (T2I) diffusion models~\cite{sd3,podell2023sdxl} are generative models that take a prompt and Gaussian noise as input and generate a realistic image aligned with the prompt.
\Cref{fig:diffusion} depicts the structure of diffusion model. The Prompt Encoder converts the prompt into embeddings used by every denoising iteration. The model then progressively predicts and removes noise, gradually transforming the noisy input into a high-quality image. The denoising component typically adopts one of two architectures: U-Net~\cite{ldm, EmoGen, dalle3, podell2023sdxl} or Diffusion Transformer (DiT)~\cite{dit, sd3, hunyuan}. U-Net combines ResNet blocks and Transformer blocks, while DiT only has Transformer blocks. The Transformer blocks apply self-attention to refine visual details and cross-attention to enhance text–image alignment. After completing all denoising steps, a decoder upsamples the output to the target resolution. Although DiT models are generally considered more powerful, U-Net–based models remain widely adopted due to their lightweight architecture \cite{xia2025modmefficientservingimage} and strong capability in condition alignment. In fact, most auto-regressive image processing \cite{chen2025blip3ofamilyfullyopen,huang2025illumeilluminatingunifiedmllm,liu2025step1xeditpracticalframeworkgeneral} services employ U-Net–based diffusion models as decoders, leveraging their efficiency to produce images with sharper details and clearer text.

Although diffusion models are capable of synthesizing high-quality images, they incur substantial overhead due to the iterative generation process. This challenge becomes even more severe in mixed-resolution serving, where diversity between the input and output shapes obstructs batch-level optimization, thereby limiting overall efficiency.

\subsection{System Optimizations for GenAI Applications}
\begin{figure}[t]
  \begin{center}
  \includegraphics[width=\linewidth]{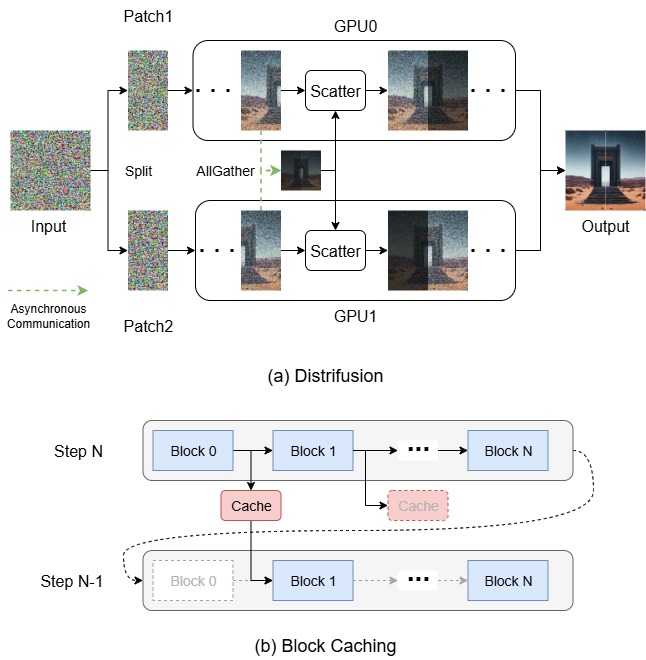}
  \end{center}
  \caption{\label{fig:distrifusion} Two T2I diffusion optimization techniques. (a) Distrifusion splits the image into multiple patches and dispatches them to different GPUs. (b) Block Caching leverages the locality, reusing block output from the previous step, and skipping the corresponding block in the current step.}
\end{figure}

To mitigate the heavy overhead in the T2I diffusion model, prior studies have primarily pursued two directions: Parallelism and Locality Exploration.

\textbf{Parallelism Exploration with Patching.} One promising direction for accelerating diffusion is to patchify images and distribute the patches evenly across multiple GPUs \cite{li2023distrifusion, fang2024pipefusionpatchlevelpipelineparallelism, xdit}.  \Cref{fig:distrifusion}(a) illustrates this approach: each image is partitioned into two patches, which are then assigned to separate GPUs for concurrent processing. To mitigate synchronization overhead, the system performs the \code{AllGather} operation asynchronously and scatters cached states to each GPU. While this strategy enhances parallelism, it still lacks support for mixed-resolution serving. Specifically, it constrains each request to a fixed number of patches determined by the number of GPUs, which hinders patch size unification, failing to batch requests with diverse resolutions.
Moreover, it only exchanges stale cross-GPU context for approximation rather than sharing up-to-date information, further exacerbating accuracy degradation.

\textbf{Locality Exploration with Caching} Another optimization technique is caching. Prior studies have observed that block outputs evolve gradually across denoising steps, enabling the reuse of previously computed results \cite{blockcache, ma2024learningtocache, deepcache}. \Cref{fig:distrifusion}(b) illustrates the central idea: the caching technique records the output of each block and selectively reuses it in subsequent steps. To keep a balance between efficiency and quality, prior works typically rely on offline profiling to determine the skipped blocks in each step. While this method reduces computation overhead, it enforces a static model configuration in which reuse decisions are predefined, limiting adaptability to dynamic input shape variations.
% Second, another array of previous works \cite{nirvana, li2023distrifusion} focusing on optimizing diffusion models ignore the importance of fully utilizing the capacity of the accelerator. For example, Distrifusion \cite{li2023distrifusion} optimizes diffusion by leveraging patch parallel method to overlap communication and computation in distributed systems. It reduces the latency but prevents diffusion from achieving benefits from batching. NIRVANA \cite{nirvana} finds that similar prompts also share similar latent states, especially for the first several steps. It attempt to eliminate such redundant steps by reusing previous latent states generated by other prompts. 
% FISEdit \cite{yu2024accelerating} and SIGE \cite{NEURIPS2022_b9603de9} replace normal convolution with sparse convolution to reduce extra computation while editing image. \todo{highlight the similarity-related technique in the above sentences, currently the information given are irrelevant}. 
% NIRVANA aims to reduce the computation of diffusion by leveraging the similarity of noise distribution, yet they overlook that such similarity would be hard to reuse when resolution changes. 

\begin{figure}[t]
\begin{minipage}[t]{0.45\linewidth}
    \centering
    \includegraphics[width=\linewidth]{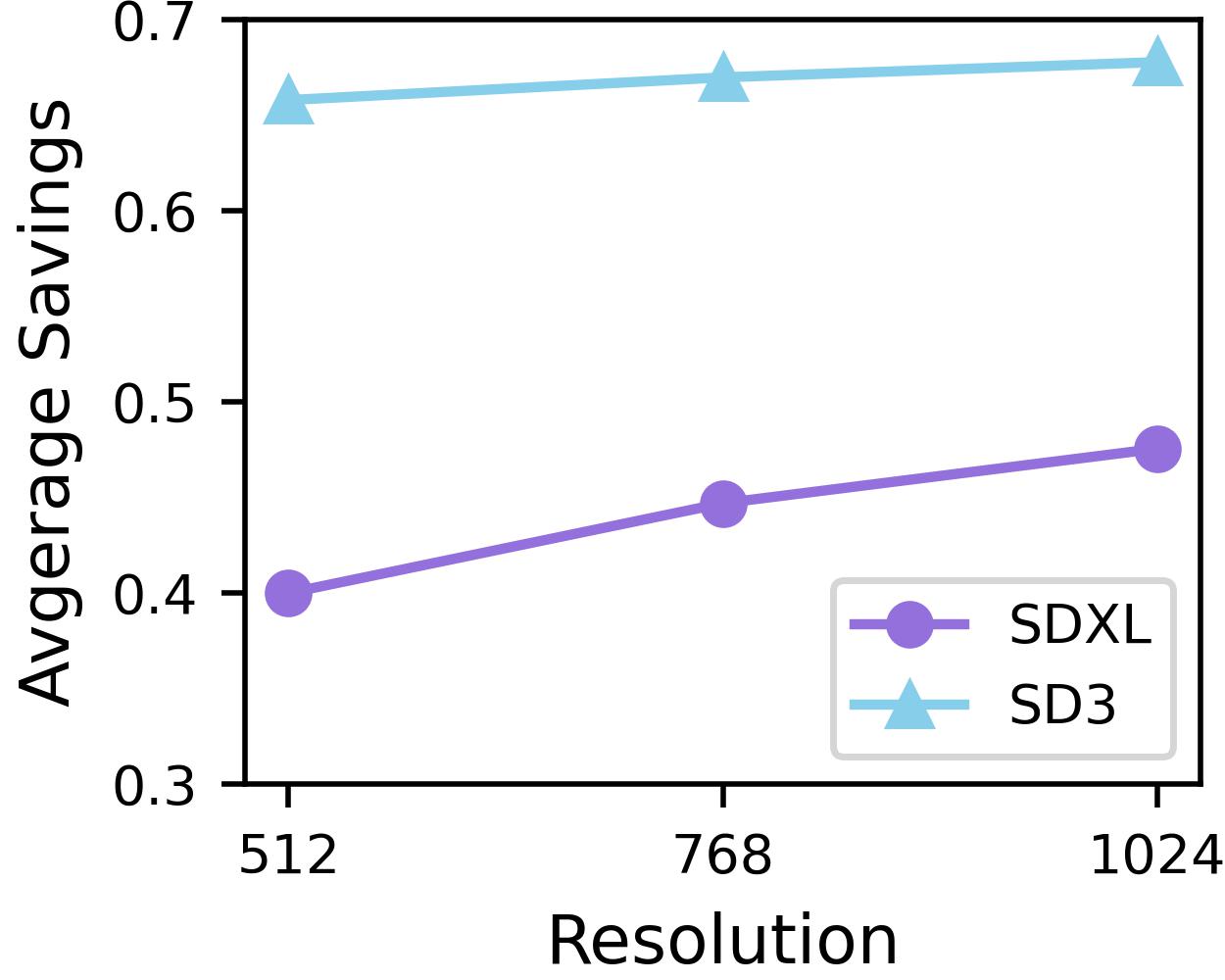}
    \caption{Average savings from skipped computations. }
    \label{fig:skip_blocks}
\end{minipage}
\hspace{0.2cm}
\begin{minipage}[t]{0.45\linewidth} 
    \centering
    \includegraphics[width=\linewidth]{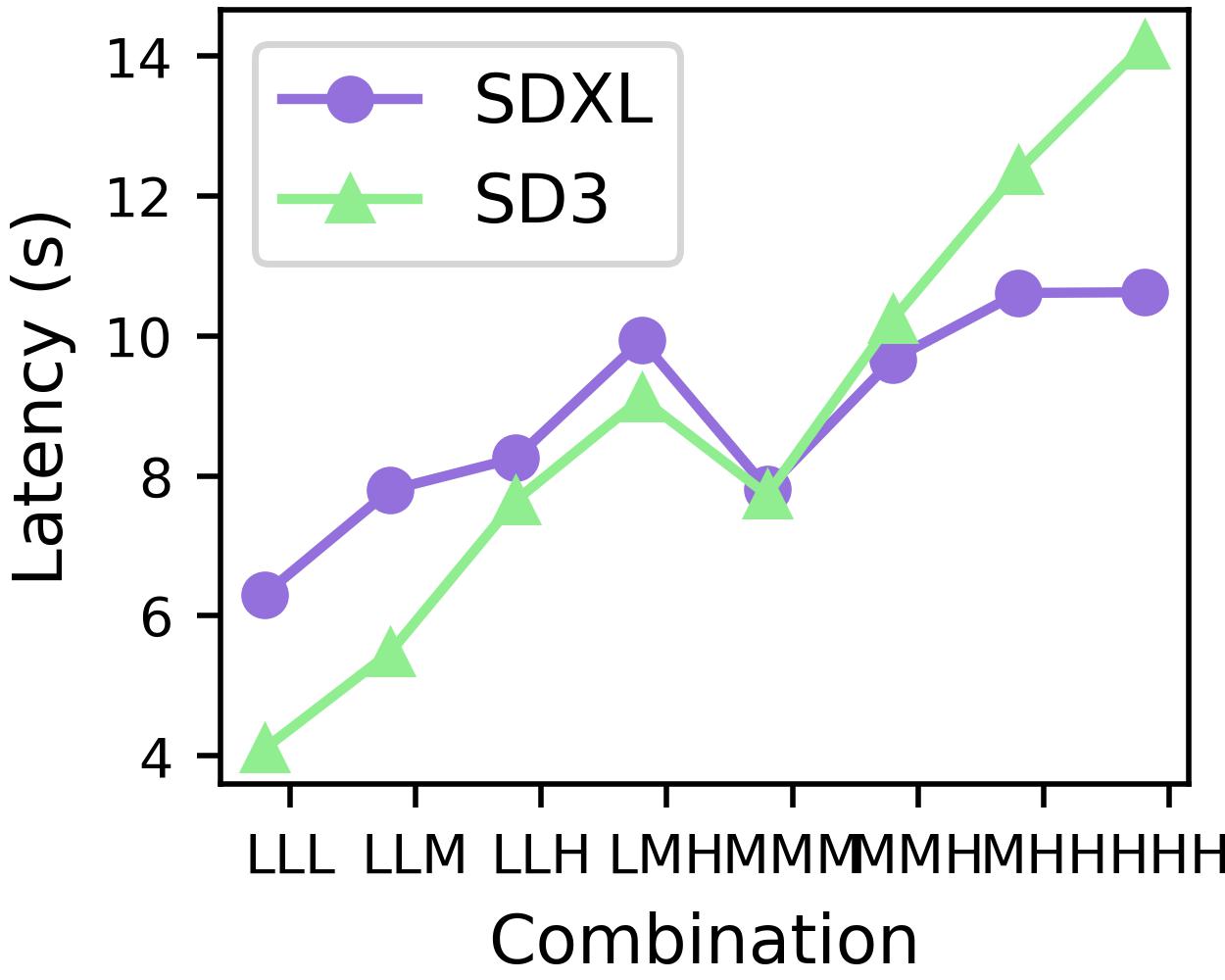}
    \caption{Latency with same batch size but the different combination.}
    \label{fig:motivation_latency}
\end{minipage}       
\end{figure}

\begin{figure*}[t]
\begin{minipage}{0.38\linewidth}
    \begin{center}
  \includegraphics[width=\linewidth]{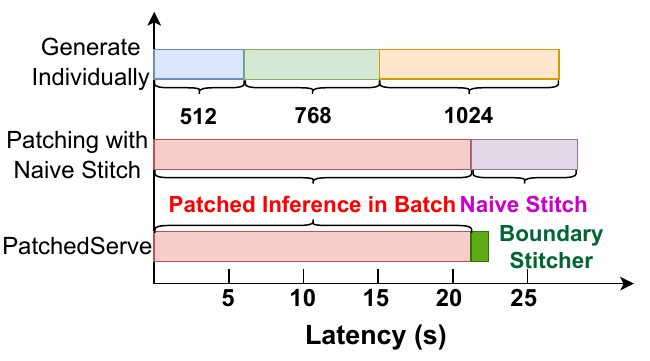}
  \end{center}
  \caption{\label{fig:stitcher} Latency Comparison with different stitching method.}
\end{minipage}       
\hspace{0.2cm}
\begin{minipage}{0.58\linewidth} 
    \centering
    \includegraphics[width=\linewidth]{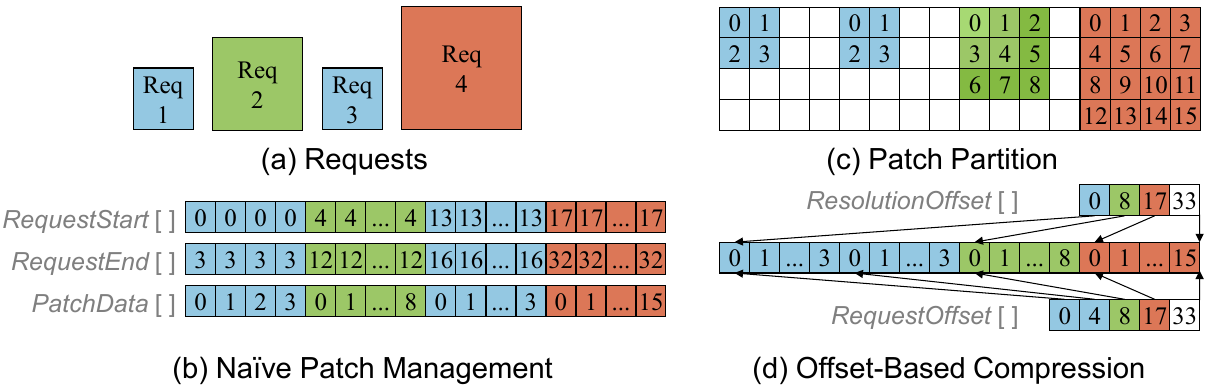}
    \caption{Patch management policy. (a) Requests with various resolutions. (b) Naive Patch Management. (c) Reorder and consider the patches as sparse arrays. (d) Exploit offset to record position.}
    \label{fig:patch_manager}
\end{minipage}   
\end{figure*}  

Beyond these two categories of optimization, several other research directions have been explored. Some studies identify inefficiencies in the iterative denoising process and propose reducing the number of denoising steps \cite{NEURIPS2022_260a14ac,lu2023dpmsolver,luo2024latent,nirvana,sun2024flexcacheflexibleapproximatecache,xia2025modmefficientservingimage}. Others focus on structural redundancy within diffusion models, advocating the introduction of sparsity to enhance efficiency \cite{Yu_Li_Fu_Miao_Cui_2024,NEURIPS2022_b9603de9,zhang2025spargeattentionaccuratetrainingfreesparse,cheng2025catpruningclusterawaretoken}. These approaches are orthogonal to our work, and advanced techniques from these directions can be seamlessly integrated into \Mname{} to further boost performance.

\section{Challenges and Motivations}

\iffalse
\begin{figure*}[!htbp]
\begin{minipage}[t]{0.19\linewidth}
    \centering
    \includegraphics[width=\linewidth]{pic/experiment/background/skip_blocks.jpg}
    \caption{Average savings from skipped computations. }
    \label{fig:skip_blocks}
\end{minipage}
\hspace{0.2cm}
\begin{minipage}[t]{0.19\linewidth} 
    \centering
    \includegraphics[width=\linewidth]{pic/experiment/background/latency.jpg}
    \caption{Latency with same batch size but the different combination.}
    \label{fig:motivation_latency}
\end{minipage}       
\hspace{0.2cm}
\begin{minipage}[t]{0.58\linewidth} 
    \centering
    \includegraphics[width=\linewidth]{pic/patch-manager.pdf}
    \caption{Patch management policy. (a) Four input requests with different resolutions. Each resolution may have a different number of requests. (b) Record latent information for all patches. (c) Reorder and consider the patches as sparse arrays. (d) Use resolution offset and Request offset to help manage patches.}
    \label{fig:patch_manager}
\end{minipage}   
\end{figure*}  
\fi

\iffalse
\begin{figure}[t]
  \begin{center}
  \includegraphics[width=\linewidth]{pic/stitcher_comparison.pdf}
  \end{center}
  \caption{\label{fig:stitcher} Latency Comparison with different stitching method.}
\end{figure}
\fi

While prior work leverages patching to hide communication overhead across multiple GPUs \cite{li2023distrifusion,fang2024pipefusionpatchlevelpipelineparallelism,xdit}, our approach aggregates multiple patches to maximize parallel throughput and incorporates caching to reduce computational overhead. However, this design introduces several key challenges: \textit{Inefficient Context Exchanging}, \textit{Mismatched Skipped Blocks}, and \textit{Explosive Combination}.

% \label{section:mip}
\iffalse
\textbf{Inefficient Patch Index: }After splitting the whole image into patches, we need some extra indices to help us identify the patches from the same request or the same resolution. {Some existing studies have proposed patching techniques \cite{li2023distrifusion,xdit}, they still suffer from suboptimal performance 
due to several reasons: 
First, they cannot effectively hide boundary exchange overhead, especially for column boundaries, which cause discontinuous memory access. Second, the diverse positions of patches and boundaries make them hard to efficiently locate and reorganize exchanged data. Unlike simple top-bottom stitching, handling multi-directional boundaries across many patches demands a custom solution to reduce context compensation overhead, which is beyond their ability. Finally, they leverage stale KV cache to hide communication overhead, which leads to accuracy loss and memory overhead.} In T2I diffusion models, there are two kinds of operations that need extra context from other patches: \textit{Convolution} and \textit{Self-Attention}. {Convolution} requires other patches' boundaries to help maintain the current patch boundary's quality, while {Self-Attention} needs to calculate the interactive relation with all the other patches from the same request (\Cref{fig:unet_kernel}). Therefore, it is essential to have efficient patch management 
to handle {\textit{Convolution}} and {\textit{Self-Attention}} in an accurate and efficient manner.
\fi

\textbf{Inefficient Context Exchanging: } Enabling patch-level parallelism in heterogeneous diffusion workloads requires locating each patch’s resolution and request metadata. However, existing patching methods~\cite{li2023distrifusion,fang2024pipefusionpatchlevelpipelineparallelism,xdit} suffer from multiple inefficiencies.
First, they fail to exchange cross-patch context within a batch since they process requests sequentially. Exchanging context within a batch leads to heterogeneous patch placements, resulting in multi-directional boundary stitching, which is more complicated than simply stitching the top boundary of one patch to the bottom of another in prior studies. \Cref{fig:stitcher} demonstrates that naively stitching will degrade performance (details in \Cref{subsubsec:stitcher}).  Second, these methods rely on stale KV caches to overlap communication latency, which introduces both accuracy degradation and memory overhead.
Consequently, efficient context completion is critical for ensuring both accuracy and scalability.

\textbf{Mismatched Skipped Blocks: }Prior work has introduced caching mechanisms to reduce redundant computation in diffusion models~\cite{blockcache, ma2024learningtocache}, typically by reusing outputs from designated blocks across denoising steps. However, these solutions rely on fixed caching patterns that fail to adapt to resolution changes, making them unsuitable for mixed-resolution serving scenarios.
To highlight this limitation, we evaluate the model at resolutions 512, 768, and 1024 across 1,000 runs with random seeds, applying the Block Caching strategy~\cite{blockcache} to measure the distribution of skipped blocks.
% Following Block Caching~\cite{blockcache}, a block is skipped if its output is similar to that from the previous step, determined by a threshold $\sigma$. Using mean squared error (MSE) with $\sigma=0.1$, we evaluate the model at resolutions 512, 768, and 1024, running each setting 1,000 times with different seeds. We then compute the average number of skipped blocks.
\Cref{fig:skip_blocks} demonstrates that the set of skipped blocks varies substantially across different resolutions, indicating the inefficiency of applying a single caching strategy uniformly.

\textbf{Explosive Combination: }Prior work often relies on offline latency profiling~\cite{INFaaS, chen2024otas, Proteus, ahmad2025diffserve} to schedule requests under SLO constraints. However, this strategy becomes infeasible in the presence of an explosive combination of resolutions. We demonstrate this by measuring latency across all resolution combinations with 3 requests, using the mixed-resolution batching method described in \Cref{sec:patch-batch}. With three resolutions, there are eight possible combinations. \Cref{fig:motivation_latency} reports the average latency for each. ``LMH'' denotes a batch with one request each at low, medium, and high resolutions. The results exhibit substantial variability--—batches composed entirely of high resolution requests can be up to 68\,\% slower than those with only low resolution requests.
To capture this variability, all combinations need profiling. If a GPU supports up to $M$ concurrent requests and there are $N$ resolutions, the number of unique combinations is $\sum_{i=1}^{M} C_{N - 1}^{i + N - 1}$, which grows rapidly with $M$ and $N$. Consequently, efficient latency estimation without full profiling is essential.

\iffalse
We observe certain patterns in the system latency behavior. First, latency is directly influenced by the number of pixels processed. As the T2I diffusion model generates more pixels, its processing speed decreases. Second, latency is also affected by the number of resolutions involved. We find that although ``LMH'' computes fewer pixels than ``MMH'', it may take longer time owing to an additional resolution ``L''. These two observations lead us to an accurate predictor which we will discuss it in \Cref{subsec:predictor}.
\fi
% As described in Figure \ref{fig:diffusion}, diffusion models are composed by three stages: one iterative denoising stage and two normal stages. Assuming there is a pair of requests ($R_i$, $R_j$), there will be two cases: (1) both of these two requests are not served, (2) one request is already executed (let's say this request is $R_i$) and the other is still waiting. For the first case, system could easily package these two requests into a batch. For the second case, previous research suggests to block $R_i$ until $R_j$ runs to the same stage as $R_i$, which leaves $R_j$ processed individually in a large proportion of inference. Another batching strategy enables add new requests after each iteration based on "selective batch" strategy. In diffusion models, since $R_i$ and $R_j$ is not start at the same time, they are destined to complete their denoising stage sequentially, resulting in separately processing Decoder stages, which miss the opportunity of batching.

% \input{sections/overview}

\section{Patched Inference with Batching}
\label{sec:patch-batch}
% We propose {\bf Patched Batching Inference} method, which highlights both model processing and batch scheduling to solve the gaps in mixed resolution diffusion serving system. 
In this section, we first introduce how we manage the patches. Next, we describe our approach for addressing missing context during inference.

\iffalse
\begin{figure}
    \centering
    \includegraphics[width=\linewidth]{pic/patch-manager.pdf}
    \caption{Patch management policy. (a) Four input requests with different resolutions. Each resolution may have a different number of requests. (b) Record latent information for all patches. (c) Reorder and consider the patches as sparse arrays. (d) Use resolution offset and Request offset to help manage patches.}
    \label{fig:patch_manager}
\end{figure}
\fi

\begin{figure*}[t]
  \begin{center}
  \includegraphics[width=\linewidth]{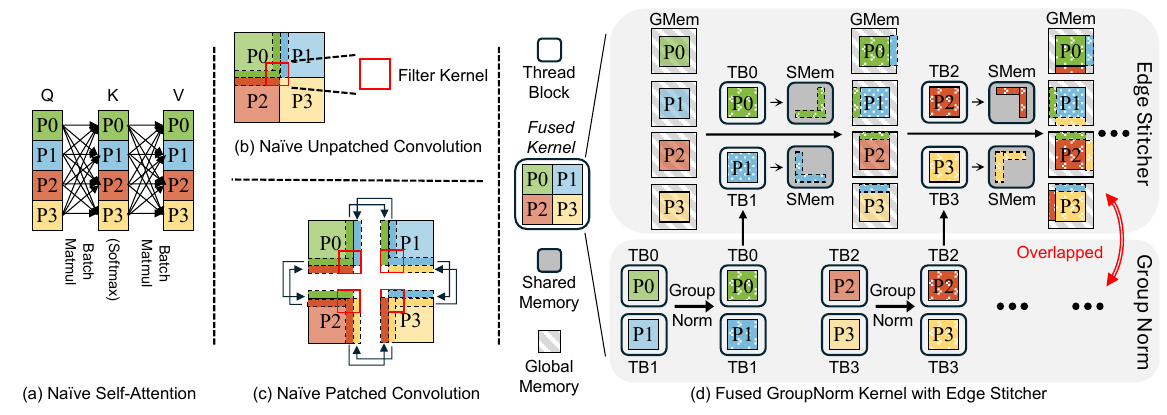}
  \end{center}
  \caption{\label{fig:unet_kernel} {Patch based operators in T2I diffusion model. (a) Naive Self-Attention operation, which computes interactions across all patches. (b) Unpatched Convolution. (c) Naive convolution after patching, which gathers boundaries from neighboring patches (d) Patch Edge Stitcher, which enables boundary stitching combined with GroupNorm.}
  % \todo{This figure looks so simple, You should give a concrete example of the process with the address and size of memory transactions.}
  }
\end{figure*}
\subsection{Compressed Sparse Patch Format}
% \todo{This section should be a new ``Section'' instead of a subsection}
 
% {\bf Why should we manage patches?} 
{Unlike prior work~\cite{li2023distrifusion, fang2024pipefusionpatchlevelpipelineparallelism,xdit}, which primarily targets distributing computation across multiple GPUs, our patching mechanism partitions images along both the height and width dimensions, selecting the patch size as the greatest common divisor of all resolutions in the corresponding dimension. This approach allows patches to be processed concurrently and leverages up-to-date data. However, operations such as \textit{Convolution} and \textit{Self-Attention} exhibit dependencies across patches, making efficient patch localization critical. }
% After splitting images, there should be a batch of patches. Although most of the operators can be processed directly, some of them need context information to maintain high quality. These operators usually need information from other patches in the same image (e.g., {Convolution} and {Self-Attention}). However, if we record the position for each patch, we have to traverse the whole position list to collect other patches from the same image. On GPU, that means every thread accesses the whole list before start the real calculation, which brings much waste of the resource.

% \textbf{Patch management strategy.} Inspired by 
Inspired by the \textit{Compressed Sparse Row} (CSR) format commonly employed for irregular data structures, we introduce a novel \textit{Compressed Sparse Patch} (CSP) representation to efficiently manage image patches (\Cref*{fig:patch_manager}). Consider four pending requests sorted by arrival time (\Cref{fig:patch_manager}(a)). Naively storing latent information for each patch individually (\Cref{fig:patch_manager}(b)) results in memory inefficiency and obstructs recovering images by resolution (\Cref{subsubsec:self-attention}). To mitigate this, we first reorder requests by resolution and treat patch data as sparse arrays (\Cref{fig:patch_manager}(c)). Owing to the locally dense distribution of patches within a single image, we only record request and resolution metadata for each patch. After compressing the sparse batch, we log the offset of the first patch per request (\Cref{fig:patch_manager}(d)) and the resolution offsets needed for \textit{Self-Attention} (details in \Cref{subsubsec:self-attention}). Each patch identifies corresponding request via the $\mathit{index}$, allowing it to traverse all associated patches by scanning from $\mathit{RequestOffset}$ $\mathit{[index]}$ to $\mathit{RequestOffset[index + 1]}$.

\iffalse
\textit{Compressed Sparse Row} (CSR) in the sparse matrix for a diverse range of irregular data structures (e.g., graphs), we design a novel \textit{Compressed Sparse Patch} (CSP) to manage the image patches efficiently (\Cref*{fig:patch_manager}). Assume there are 4 requests in waiting list sorted by their arrival time (\Cref{fig:patch_manager}(a)). If we simply record the latent information for every patch (\Cref{fig:patch_manager}(b)), not only do we waste memory space, but also miss the opportunity to gather all requests with the same resolution for some operators (like Self-Attention in \Cref{subsubsec:self-attention}). Therefore, we first reorder the requests by resolution and then treat them as sparse arrays (\Cref{fig:patch_manager}(c)). 
For each sparse array, non-zero patches are always at the left-top position, therefore we don't need to log the location in images for patches but only record the request and resolution information. After compressing the batch of sparse images to a list of patches, we first note the offset of the first patch in every image (\Cref{fig:patch_manager}(d)). We also write down the resolution offset for {Self-Attention} (We will discuss it in \Cref{subsubsec:self-attention}). Each thread holds the $\mathit{index}$ of request offset which helps them identify the request the thread is processing. When one operator needs the context information, GPU can fetch other patches by simply traversing $\mathit{Request Offset[index]}$ to $\mathit{Request Offset[index + 1]}$.
\fi

\subsection{Patch-Tailored Diffusion Operators}
\label{subsec:patch_kernel}
\label{subsubsec:self-attention}

Most operators in diffusion models, including Linear, FeedForward, and Cross Attention, operate independently for each pixel and can thus be considered ``pixel-wise'' operators. In contrast, certain operators require context from other patches; otherwise, the output would be fragmented.
Typically, two operators require context information for consistent image generation: \textit{Self-Attention} and \textit{Convolution}.

% \begin{enumerate}
{\bf Patch-based Self-Attention Module:} Self-Attention aggregates each pixel with all other pixels within the same image. It operates on three inputs: query, key, and value. The query token computes with all keys, applies a Softmax, and performs another dot product with all values. Although this process is straightforward for unpatched image, it becomes significantly more complex with patched image.
As illustrated in \Cref{fig:unet_kernel}(a), accurate Self-Attention computation in patched configuration requires each patch to compute with all other patches, forming a Cartesian product of interactions. 
This complexity makes it difficult to implement efficient GPU kernels. To address this, we reconstruct patches back into the full image before executing Self-Attention. To further enable parallel execution, we group requests by resolution, which can be achieved simply and efficiently by exploiting CSP format, to achieve efficient batched attention.

\iffalse
By leveraging CSP format, \Mname{} quickly locates and aggregates patches from requests with the same resolution, allowing larger batch sizes and improving operator throughput compared to handling requests independently.
\fi

% \item 
{\bf Patch-based Convolution:} Convolution operator applies a small kernel to aggregate features from neighboring pixels. The kernel size ranges from 1 to 3 in T2I diffusion models~\cite{podell2023sdxl, pmlr-v162-nichol22a, ldm}. When the kernel size exceeds 1, computation requires adjacent pixel values, introducing cross-patch dependencies. As illustrated in \Cref{fig:unet_kernel} (b), unpatched convolution proceeds seamlessly across the image, whereas patched convolution encounters boundary issues. For example, processing the bottom-right corner of $P_1$ requires boundary data from $P_2$ and $P_3$, while $P_1$ simultaneously provides its boundaries to these patches. These dependencies are complicated by two forms of diversity: (a) \textit{Direction Diversity}. Patches must stitch both row and column boundaries. Row boundaries align with memory layout, but column stitching incurs irregular memory access. (b) \textit{Position Diversity}. Each patch has different neighbor positions, for example, $P_1$ stitches on the right and bottom, while $P_4$ stitches on the top and left (\Cref{fig:unet_kernel}(c)). 
To enhance parallelism, we record each patch’s adjacent neighbors during splitting and pad with 0 when a neighbor is absent. This metadata supports uniform and efficient boundary stitching across all patches. Additionally, we employ a tailored stitcher to overlap memory movement overhead arising from these diversities (\Cref{subsubsec:stitcher}).

\subsection{Patch Edge Stitcher}
\label{subsubsec:stitcher}
\iffalse
We have outlined how to preserve quality in patch-based execution. However, for convolution, naïve edge stitching implemented by concatenation operation introduces significant overhead due to fragmented access patterns and costly pixel fetching across arbitrary patch boundaries.
\fi
We conduct an experiment to quantify the overhead of stitching. Each resolution is assigned four requests in our evaluation. \Cref{fig:stitcher} demonstrates that naive stitching (fetching all required boundaries and concatenating them with target patches) offsets the performance gains of patch-level parallelism, highlighting the necessity for an efficient stitcher.
We propose a lightweight patch edge stitcher that reduces memory footprint. The key observation is that convolution in diffusion models typically follow a \code{GroupNorm} operation \cite{podell2023sdxl, ldm, pmlr-v162-nichol22a}. Therefore, we fuse the stitching operation into the \code{GroupNorm} kernel. As illustrated in \Cref{fig:unet_kernel} (d), we relocate boundary pixels during normalization operations, mitigating redundant memory footprint.
Specifically, each GPU thread block (TB) normalizes one patch and checks whether its boundary pixels are required by other patches. Such dependencies are prepared during patch splitting. The boundary pixels required by other patches are temporarily saved in shared memory. After completing all normalizations in the current TB, the TB then locate the target patches of those boundaries and write them back to global memory. 
This design overlaps edge stitching with other normalizations, ensuring the convolution's accuracy without additional synchronization. The result in \Cref{fig:stitcher} demonstrates the minimal overhead from our stitcher, allowing patched execution to achieve its intended parallel throughput.

\section{Exploiting Patch-Level Locality}
\label{sec:patch-cache}

\begin{figure}[t]
  \begin{center}
  \includegraphics[width=\linewidth]{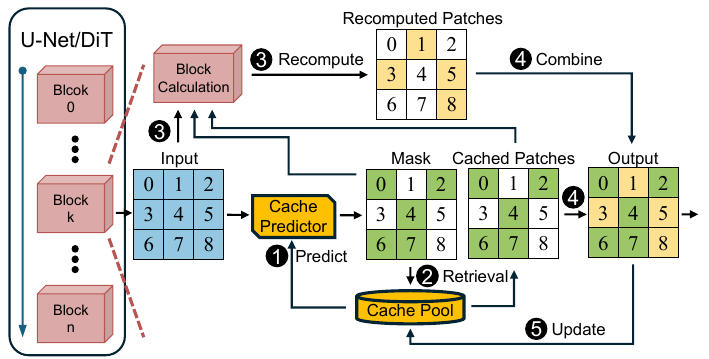}
  \end{center}
  \caption{\label{fig:patched_cache} {Patch-level cache reuse system overview. }
  % \todo{This figure looks so simple, You should give a concrete example of the process with the address and size of memory transactions.}
  }
\end{figure}

To maintain image quality and reduce computational overhead, we propose a patch-level cache reuse strategy. We determine whether to reuse the cache dynamically before each block in each step, ensuring that only patches with significant deviations from cached data are recomputed. To minimize the overhead from cache operations, we coalesce multiple cache operations to process them simultaneously.
% {To enhance the image quality and save computations, we propose to reuse cache at patch level. We make the decision whether to reuse the cache before each block, and only calculate the patches that indicate huge difference from the cache. To query the cache online, we also design to manage cache in batch .}

\subsection{Patch-aware Caching Strategy}
\Cref{fig:patched_cache} depicts the workflow of patch-level caching, which is applied before every blocks in T2I diffusion models.
When a new input comes, \textit{Cache Reuse Predictor} (Later discussed in \Cref{sec:impl}) compares input and cache from the previous step (\scalebox{0.8}{\circled{1}}).  The predictor generates a mask determining the reusability for each patch. (\scalebox{0.8}{\circled{2}}).
The input and generated mask are subsequently forwarded to the current block (\scalebox{0.8}{\circled{3}}). For pixel-wise operators, recomputing only the unmasked patches is sufficient. However, as discussed in \Cref{subsec:patch_kernel}, certain operations rely on features from other patches to preserve quality. If masked patch values are directly used as inputs for such operations,  the result may mismatch in shapes or output with significant errors. Fortunately, prior studies \cite{li2023distrifusion,xdit} observe that the outputs of operators from adjacent steps are sufficiently similar, allowing us to reuse the results from the previous step to fill the masked patches.
After the block execution, part of the output is imprecise due to the masked processing pattern. Therefore, we use the mask again to replace the masked patches with cache, which is generated from the last step (\scalebox{0.8}{\circled{4}}). Finally, the system updates the input and output of this block for the next step (\scalebox{0.8}{\circled{5}}).

\subsection{Batching Patches in Cache}

Since we should access cache every single operation to load or save data, it is obvious that cache management affects the extent of benefit from caching. In SD3~\cite{sd3}, it takes 40 to 50 ms to process one step with 24 blocks, which means we have to use less than 2 ms to complete all the cache operations for a single block, otherwise we cannot gain any profits even if all blocks can be skipped. To achieve this, the cache system should support three fundamental operators: \code{query}, \code{delete}, and \code{update/insert}.

\begin{figure}[t]
% \begin{minipage}[t]{0.44\linewidth} 
    \centering
    \includegraphics[width=\linewidth]{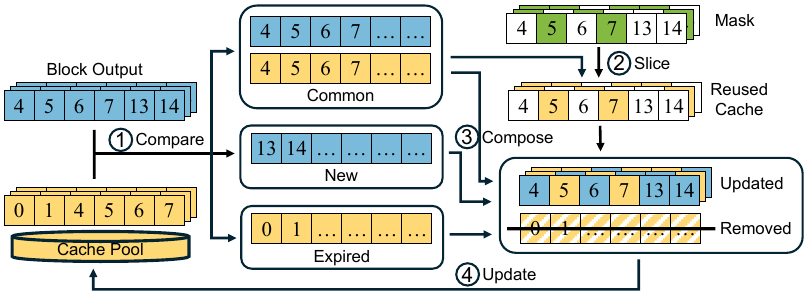}
    \caption{Batching Patches in Cache.}
    \label{fig:cache_management}
% \end{minipage}       
% \hspace{0.2cm}
% \begin{minipage}[t]{0.54\linewidth} 
%     \centering
%     \includegraphics[width=\linewidth]{pic/MixFusion.pdf}
%     \caption{Overview of \Mname.}
%     \label{fig:arch}
% \end{minipage}   
\end{figure}  

To efficiently manage the cache system, we adopt a batching strategy to amortize the cache overhead by processing them concurrently. We employ a \code{map} to store cache data, where each patch is assigned a unique identifier, and each block maintains an independent cache. \Cref{fig:cache_management} depicts the overall design. When a block needs to generate a mask or compute a masked output, it submits indices which consist of each patch's unique ID along with intermediate results to the cache system as input. By comparing these indices with the entries stored in the cache, the system identifies three distinct sets (\scalebox{0.8}{\whitecircled{1}}): 
\textit{Common Set:} IDs present in both the cache and the input indices. When receiving new data, the cache system verifies whether the cache should replace the masked patched in the input (\scalebox{0.8}{\whitecircled{2}}).
\textit{New Set:} IDs present only in the input indices, which will be inserted into the cache. Unmasked regions are recomputed and then updated in the cache, while the New Set provides missing indices and coalesces them for updating (\scalebox{0.8}{\whitecircled{3}}).
\textit{Expired Set:} IDs present only in the cache. Since preemption is not allowed in \Mname, each patch will stay on GPU until it finishes. Once the cache system detects IDs that is only in the cache system, it concludes that the corresponding patch has exited. Finally, the system removes expired entries from the Expired Set (\scalebox{0.8}{\whitecircled{4}}). By coalescing these operations, the cache system enables scalable patch reuse in parallel execution.

\section{SLO-Aware Scheduler}
\label{sec:schedule}

\begin{algorithm} [t] \small
\caption{\Mname~schedule algorithm}%算法名字
	\label{algorithm:schedule}
	\LinesNumbered %要求显示行号
	\KwIn{$wait\_queue, act\_queue$}
        \KwOut{$active\_queue$}
  
  \While{\mbox{True}}{
    $cur\_task \leftarrow get\_least\_slack\_task(wait\_queue)$\\

    $act\_task \leftarrow get\_least\_slack\_task(act\_queue)$\\
  $pred\_latency \leftarrow predictor(cur\_task, active\_queue)$\\
  \textcolor{blue} {/*SLO Violation Analyze*/}\\
  \If {\mbox{time\_out(cur\_task, pred\_latency)}} {

    $discards(cur\_task)$\\

    continue

  }
  \textcolor{blue} {/*Schedule Mode Decision*/}\\
  \If {\mbox{switch\_mode(cur\_task, pred\_latency)}} {
    $cur\_task \leftarrow update\_task()$\\
    $pred\_latency \leftarrow update\_latency(cur\_task, act\_queue)$\\
  }

  \textcolor{blue} {/*Schedulability test*/}\\
  \If {\mbox{time\_out(act\_task, pred\_latency)}} {
    break
  } \Else{
    act\_queue.enqueue(cur\_task)
  }

  % $wait\_heap, active\_heap \leftarrow regenerate\_heap(cur\_task)$

  }
  % return $\mathit{active\_queue}$
\end{algorithm}

\subsection{Mixed-Resolution Throughput Analyzer}
\label{subsubsec:analyzer}
Admitting a new request into the current batch requires careful consideration. For instance, when the Schedule Decider considers admitting a new task into the current batch, the task introduces additional overhead. While this may improve overall throughput, it can also increase batch latency, risking SLO violations for some tasks. Such complex trade-off emphasizes the importance of accurate latency prediction.
To make efficient scheduling decisions, \Mname~employs a \textit{Throughput Analyzer} that forecasts the future latency of the potential batch. Conventional systems often rely on offline profiling to estimate model execution latency \cite{INFaaS, chen2024otas, Proteus}, while such solution performs bad on mixed-resolution T2I diffusion serving system. This scenario introduces a significantly larger set of possible task combinations, making exhaustive offline profiling infeasible. Moreover, since \Mname~combines requests into a single batch, the actual latency is typically less than the sum of per-task latencies. Overestimating this latency discourages admitting new tasks, ultimately reducing system throughput.

Based on these considerations, the Throughput Analyzer employs Multilayer Perceptron (MLP) for latency prediction. The MLP model takes three inputs: the task number for each resolution, the ongoing resolution number, and the total patch number. We generate 200 diverse resolution combinations and evaluate their latencies as the dataset, where 80\,\% of which is the train set and the remaining is the eval set. The MLP model achieves high prediction accuracy, with errors of less than 3.7\,\% compared to the actual latency, indicating negligible runtime overhead.

\subsection{SLO-based Scheduling Algorithm}
\label{subsec:scheduler}

In mixed-resolution settings, each request may have distinct resolution and SLO requirements, and the system’s ability to split requests into arbitrary patch sizes further complicates the decision of an optimal strategy, which necessitates an effective scheduling algorithm.
Suppose there are $M$ different resolutions, with $\mathit{N_i}, i \in M$ requests in the waiting queue. Moreover, each request has distinct urgency, leading to an exponentially growing search space.
\iffalse
\Mname{} can select from $P_i$ choices, where each resolution combination leads to one option. This results in an enormous search space with complexity $\mathit{O(MNP)}$.
\fi
Completing such an exhaustive schedule selection within a single scheduling period is therefore highly challenging.

To address this challenge, we introduce a heuristic scheduling algorithm that reduces scheduling overhead to a practical level while preserving high SLO satisfaction (Algorithm~\ref{algorithm:schedule}). First, we define the slack score for request i as: 

$$ Slack_i = \frac{DDL_i - C_i - P_i}{SA_i}  $$

Here, \textit{$DDL_i$} and \textit{$SA_i$} denote the SLO constraint and the standalone model latency of request $i$, respectively. $C_i$ represents the time consuming since request $i$ arrived, and \textit{$P_i$} is the predicted time of the remaining stages. The slack score quantifies request urgency, with lower scores indicating higher priority for earlier execution. This scheduling procedure can be performed in parallel with the denoising computation. 

The scheduler is designed to balance throughput and SLO requirements, providing an efficient and systematic approach. The scheduler chooses either the most urgent request to prevent starvation or the one that maximizes throughput improvement for the current batch.  
If the most urgent request still has a relatively relaxed slack, the scheduler switches to a throughput-optimized mode and selects the next candidate (lines 11-14). This selection process continues until no additional requests can be admitted without violating SLO constraints (lines 16–18).
If a candidate cannot meet its deadline even when processed immediately, it is discarded, consistent with prior approaches (lines 6-9) \cite{AlpaServe, cd-msa}.
\iffalse
Additionally, if adding a new task to the active queue risks causing a timeout for any current task, the enqueue operation is halted (lines 9-11). 
\fi

% \todo{I would suggest add the simple comments at the key step in the algorithm}

\section{Implementation}
\label{sec:impl}

We implement \Mname~with 12.5K line of codes in Python and C++/CUDA based on PyTorch~\cite{NEURIPS2019_bdbca288} and following the system design principles of vLLM~\cite{vllm}. Stable Diffusion~\cite{diffusers} is ported into our framework and decomposed into three stages: \textit{Preparation}, \textit{Denoising}, and \textit{Postprocessing} to implement both baseline and \Mname~more flexibly. 
Common components of the sampler are reorganized to enable batch denoising across variable denoising steps. We further integrate xformers~\cite{xFormers2022} to accelerate both baseline and \Mname. 
For prediction tasks, we leverage Scikit-learn~\cite{sklearn} to train the MLP-based Throughput Analyzer and cuML~\cite{cuml} for cache predictor. The cache predictor employs a Random Forest Classifier on the GPU to achieve high performance while Throughput Analyzer is on CPU to hide the scheduling overhead. We collect input-output similarity metrics (MSE) across all blocks and timesteps for 1K inference requests, which are then used to train cache predictor.

\section{Experiment}
\label{section:experiment}
{\bf Platform:} We conduct our experiments on a server equipped with an H100-80GB GPU and an AMD EPYC 9534 64-core CPU. The software stack consists of Ubuntu 18.04, CUDA 12.3, and PyTorch 2.2.2.

\begin{figure}[t]
  \begin{center}
    \includegraphics[width=\linewidth]{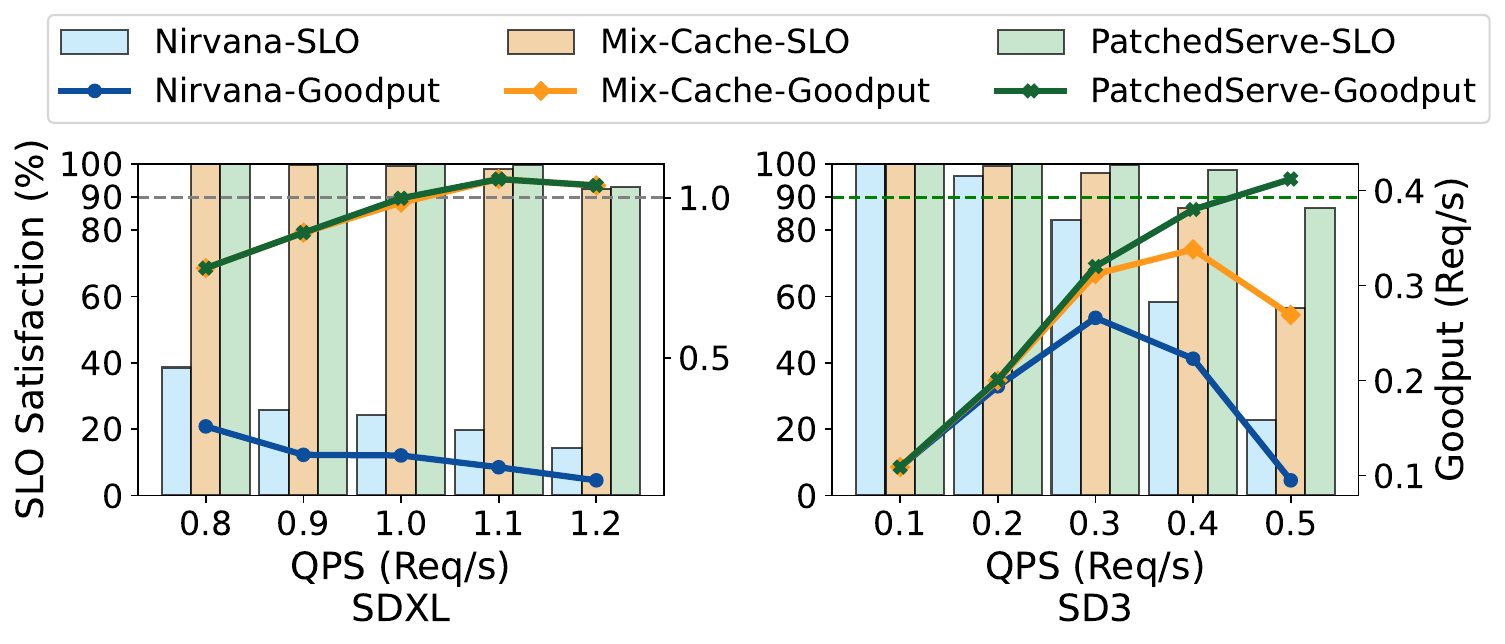}
  \end{center}
  \caption{\label{fig:performance} End-to-End SLO satisfaction ratio.}
\end{figure}

{\bf Models:} We evaluate our system using Stable Diffusion 3~\cite{sd3} and Stable Diffusion XL~\cite{podell2023sdxl}. By default, we process 50 steps for both SD3 and SDXL. Following common practice~\cite{ldm, Yu_Li_Fu_Miao_Cui_2024, nirvana, midjourney}, we adopt three widely used resolutions, $512\times512$, $768\times768$, and $1024\times1024$, noted as \textit{Low}, \textit{Medium}, and \textit{High}, as the baseline settings. Unless otherwise noted, all experiments are conducted with float16 precision.

{\bf Baseline:} We compare \Mname~with the following systems: (1) {\bf \textit{NIRVANA} \cite{nirvana}:} The state-of-the-art T2I diffusion serving system. We also incorporate the ORCA~\cite{280922} to enhance its batch size.  (2)  {\bf \textit{Distrifusion}~\cite{li2023distrifusion}:} A distributed parallel inference engine for diffusion. We only evaluate it on multiple GPUs.  (3) {\bf \textit{Mixed-Cache}:} A variant of our approach that replaces our SLO-aware scheduler with an FCFS scheduler while enabling batching.  All scheduling methods have a maximum batch size of 12 due to memory limits.

{\bf Workload:} We use COCO~\cite{chen2015microsoft} and DiffusionDB~\cite{wang-etal-2023-diffusiondb} to evaluate how much \Mname{} affects the image quality. From each dataset, we sample 5K text–image pairs to construct evaluation subsets for quality measurement. We generate input streams following a Poisson distribution, consistent with prior work~\cite{Clockwork}, where all resolutions contribute equally to the whole workload. Following the convention in Clockwork~\cite{Clockwork}, we configure the SLO requirement as 5$\times$ the execution latency for each resolution setting. 

\iffalse
\textbf{Metrics:} We evaluate performance using two primary metrics: SLO satisfaction, which quantifies the percentage of requests meeting the defined objectives, and Goodput, which measures the rate of requests completed before deadline. 
\fi
% We will display how much \Mname~outperforms compared to SOTA diffusion serving system and patch-based diffusion inference framework.

\subsection{End-to-End Performance}

We first display the end-to-end performance of \Mname~with default environment settings.

% \begin{figure}[t]
%   \centering
%     \subfloat[Gamma]{\includegraphics[width=0.5\linewidth]{pic/experiment/avg_batch_gamma.jpg}
%     \label{fig:batch-size-gamma}}
%     \subfloat[Uniform]{\includegraphics[width=0.5\linewidth]{pic/experiment/avg_batch_uniform.jpg}
%     \label{fig:batch-size-uniform}}
%   \caption{\label{fig:batch} Average Batch Size in Different Request Distribution}
% \end{figure}

% \begin{figure}[t]
%   % \centering
%     \centering
%         \subfloat[]{
%                 \includegraphics[width=0.5\linewidth]{pic/experiment/slo_with_distrib.jpg} \label{fig:slo-distrib}
%             }
%         \subfloat[]{
%                 \includegraphics[width=0.5\linewidth]{pic/experiment/goodput_with_distrib.jpg}  \label{fig:goodput-distrib}
%             }
%   \caption{\label{fig:distrib} SLO Satisfaction and Goodput trend with the change of different distribution of tasks' resolution. Small-lead means that in this scenario, tasks with small resolution takes 50\% part of the whole requests and the other two resolution of tasks takes 25\% part of the whole requests. FCFS runs OOM while executing Large-lead distribution of SDXL.}
% \end{figure}

{\bf Performance:} We first evaluate \Mname~with various QPS (Query Per Second) for both models. We only set diffusiondb \cite{wang-etal-2023-diffusiondb} as the default database in performance evaluation since \Mname{}'s performance is not affected by prompts.
\Cref{fig:performance} presents the end-to-end SLO and goodput results. Compared to NIRVANA, \Mname{} achieves 30.1\,\% higher SLO satisfaction on average while maintaining over 90\,\% SLO. The improvement is particularly pronounced on SDXL, where larger batch sizes provide greater performance gains. Specifically, SLO satisfaction drops sharply for SD3 as QPS increases, while it remains largely stable for SDXL. This is because latency gaps across resolutions are less pronounced in SDXL: generating a high resolution image takes only 1.3$\times$ the time of generating a low one, whereas SD3 requires over 2.4$\times$ longer. The larger variance in SD3 limits the arriving rate of large resolution requests, but leaves more room for scheduling optimization. This is also the reason why \Mname{} outperforms Mixed-Cache more in SD3, demonstrating the effectiveness of our scheduling algorithm. In conclusion, \Mname{} achieves 5.33$\times$ and 1.06$\times$ higher goodput when achieving 90\,\% \cite{distserve} SLO (green line in \Cref{fig:performance}) over NIRVANA and Mixed-Cache, respectively.

\iffalse
To achieve 99\,\% \cite{cd-msa, AlpaServe}, 95\,\% \cite{10.1145/3698038.3698529, cheng2024slooptimizedllmservingautomatic, clover}, and 90\,\% \cite{distserve} of the requests satisfying their SLO, \Mname{} supports 1.5$\times$, 1.12$\times$, 1.1$\times$ tighter workload on average.
\fi
% For average latency, \Mname{} demonstrates over 10\,\% improvement than Mixed-Patch owing to its preference of tighter SLO requirement requests, which are usually shorter requests in this case. NIRVANA demonstrates minimal variation with increases in QPS, indicating its inability to benefit from combining larger batch sizes at the initial stages. In contrast, Mixed-Patch and \Mname{} exhibit improved throughput under heavier workloads, suggesting greater potential for leveraging larger batch sizes.

\begin{table}[t] \small
\setlength{\tabcolsep}{4pt}
	\caption{\label{table:quality} Quality Score comparison.}
	\begin{center}
    \begin{tabular}{cccccc}
        \toprule
         Model & Method & \multicolumn{2}{c}{SDXL} & \multicolumn{2}{c}{SD3} \\
        \midrule
          & & \makecell[c]{COCO} & \makecell[c]{diffusiondb} & \makecell[c]{COCO } & \makecell[c]{diffusiondb}\\
        \midrule
        \multirow{2}{*}{\centering CLIP ($\uparrow$)} & Original & 14.92 & 16.24 & 14.79 & 16.65 \\
        & \Mname & \textbf{15.43} & \textbf{16.62} & \textbf{15.13} & \textbf{17.06} \\
        \midrule
        \multirow{2}{*}{\centering FID ($\downarrow$)} & Original & 31.92 & 35.56 & 28.94 & \textbf{32.38} \\
        & \Mname & {\bf 28.85} & {\bf 33.42} & {\bf 26.56} & 38.01 \\
        \bottomrule
    \end{tabular}
	\end{center}
\end{table}

{\bf Quality:} Table~\ref{table:quality} reports the CLIP \cite{pmlr-v139-radford21a} and FID \cite{NIPS2017_8a1d6947} scores for both datasets and both models. The CLIP score measures alignment between generated images and input prompts, with higher values indicating stronger semantic consistency, while the FID score evaluates distance between the generated images and the datasets, with a lower value representing closer to the dataset. \Mname~achieves CLIP and FID scores comparable to the original models, demonstrating that our system introduces only negligible quality degradation.

% Tail latency is the 99\%th requests' latency which can influence the overall performance in the whole requests queue. This experiment only run on gamma distribution to better reflect the affect of burst requests. As we can see, \Mname~decrease tail latency by 22.9\% and 35.9\% compared to the second best method on two models. For specific resolutions of tasks, ORCA-greedy displays better performance when the task wants a image with larger resolutions than small one, owing to its greedy strategy which prefers to continue processing large resolution tasks since large resolution of tasks stays more time on workers and more likely to receive the next task with the same resolution. This can also explain why ORCA-greedy shows worse performance on SDXL than on SD1.5. It would take more time using SDXL, which leaves more time for ORCA-greedy to find the next request with the same resolution as the ongoing one. Thus small tasks will have to wait until all the larger tasks are finished. FCFS and FCFS-MIXED show no preference with resolution changes because they are not resolution-aware. With resolution get large, the tail latency of \Mname~tends to increase. That's because task with larger resolution has a more slack deadline and it can wait more time than other tasks. 

\subsection{Sensitivity Study}

\iffalse
\begin{figure*}[t]
\begin{minipage}[t]{0.64\linewidth} 
    \centering
    \includegraphics[width=\linewidth]{pic/experiment/sensitivity_gpu_num_sdxl.jpg}
    \caption{\label{fig:multi-gpus-sdxl} SDXL End-to-End SLO change with different number of GPUs.}
    % \label{fig:cache_management}
\end{minipage}       
\hspace{0.2cm}
\begin{minipage}[t]{0.34\linewidth} 
    \centering
    \includegraphics[width=\linewidth]{pic/experiment/patch_size_throughput.jpg}
    \caption{\label{fig:patch-size-analyze} Average latency changing with patch sizes.}
\end{minipage}   
\end{figure*}  
\fi

% \begin{figure}[t]
%   % \centering
%   \begin{center}
%   \includegraphics[width=\linewidth]{pic/experiment/sensitivity_gpu_num_sd3.jpg}
%   \end{center}
%   \caption{\label{fig:multi-gpus-sd3} SD3 end-to-end SLO change with different number of GPUs.}
% \end{figure}

\begin{figure}[t]
  % \centering
  \begin{center}
  \includegraphics[width=\linewidth]{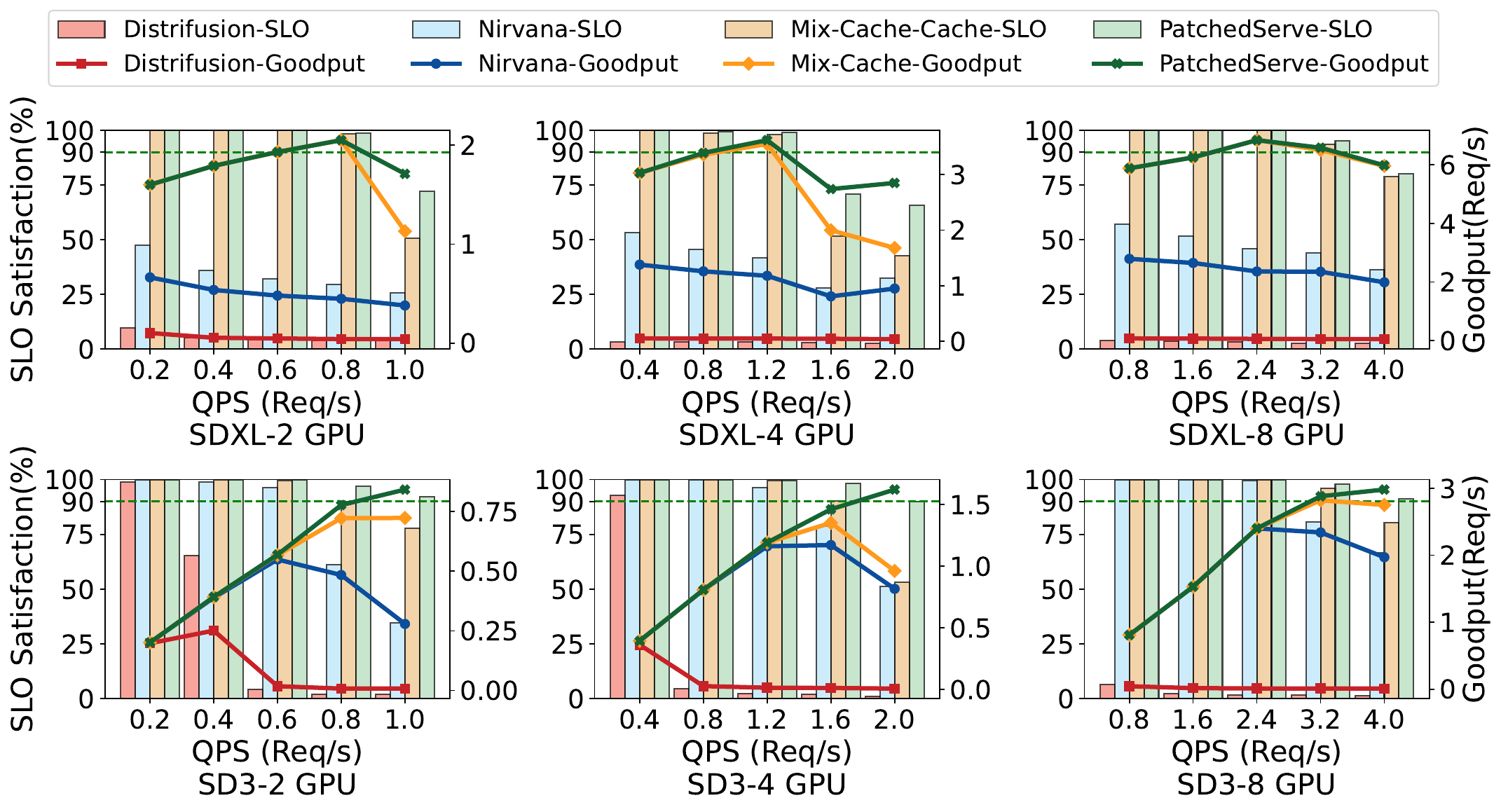}
  \end{center}
  \caption{\label{fig:multi-gpus} SDXL and SD3 end-to-end SLO change with different number of GPUs.}
\end{figure}

\textbf{Scalability. }
We further extend our evaluation to 2, 4, and 8 H100 GPUs within a single node to assess scalability. For all methods except Distrifusion, we employ data parallelism to improve load balancing. Upon the arrival of a new request, we select the GPU that has the lowest workload and dispatch the request accordingly. 
% \Cref{fig:multi-gpus-sdxl} and \Cref{fig:multi-gpus-sd3} 
\Cref{fig:multi-gpus}
demonstrates that \Mname~achieves the highest SLO satisfaction across all configurations. In contrast, NIRVANA and Distrifusion exhibit opposing behaviors: NIRVANA performs relatively better under heavy workloads, while Distrifusion only maintains high SLO satisfaction under light workloads. NIRVANA employs ORCA to form batches, thereby increasing the likelihood of incorporating additional large-resolution requests due to longer execution time, leading to stable SLO satisfaction under heavy workloads. Distrifusion, however, processes requests sequentially, which offers lower latency but fails to sustain high throughput under heavy workloads.  

\begin{figure}[t]
  \begin{center}
    \includegraphics[width=\linewidth]{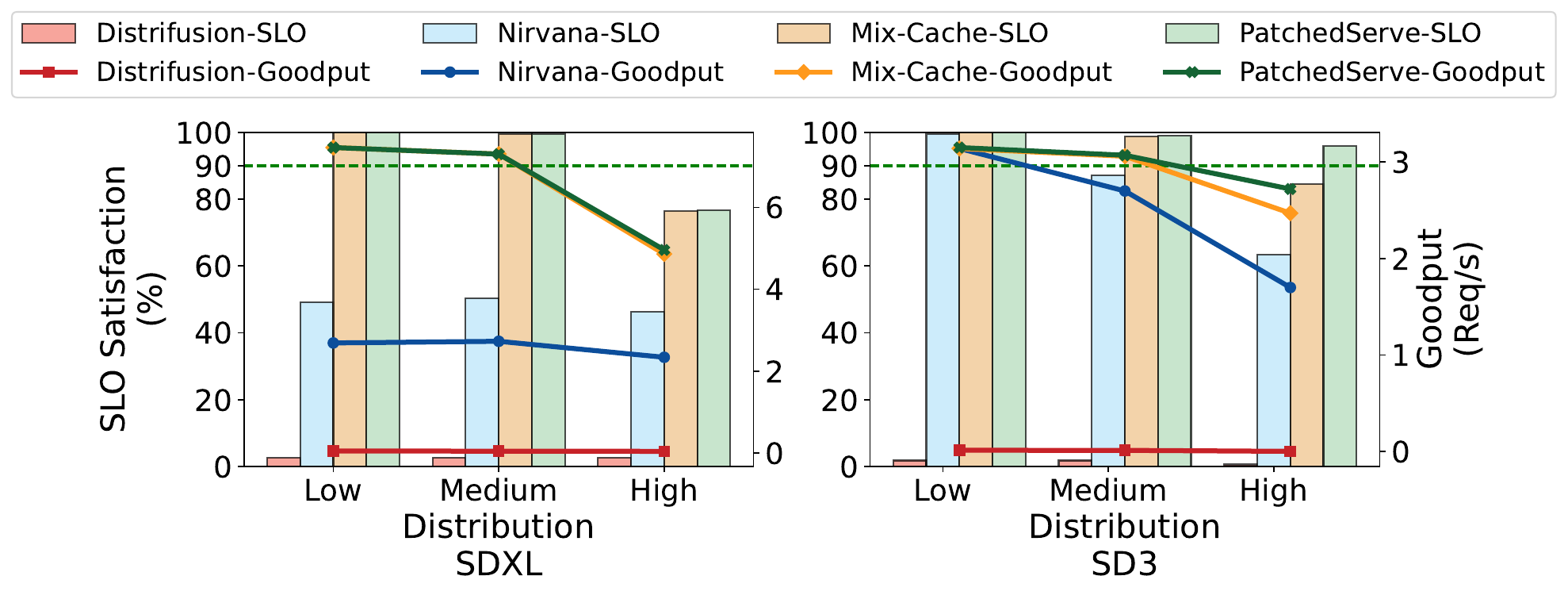}
  \end{center}
  \caption{\label{fig:sensitivity-distribution} Performance under various distribution.}
\end{figure}

\noindent \textbf{Workflow Efficiency. } We further evaluate scenarios where one resolution dominates the workload (50\,\%) while the other two share the remaining 50\,\%. We conduct an experiment with QPS of 8.8 req/s for SDXL and 3.2 req/s for SD3 on 8 GPUs. AS \Cref{fig:sensitivity-distribution} shows that \Mname{} demonstrates the highest SLO satisfaction and goodput all the time. Although \Mname{} only outperforms a little over Mixed-Cache when serving SDXL model due to similar SLO constraints, it manifests up to 11.4\,\% SLO improvement and 1.1$\times$ higher goodput than Mixed-Cache with over 90\,\% SLO on SD3. Moreover, \Mname{} demonstrates superior performance when high-resolution requests dominate the workload, highlighting its scalability under various scenarios.

\noindent \textbf{SLO Scale.} We further evaluate \Mname{} under different SLO scales to examine its behavior across varying constraints. We adopt the same configurations as in the workflow study. Notably, NIRVANA outperforms Mixed-Cache when the SLO is set to 3$\times$ the baseline latency on SD3. This advantage comes from the ORCA scheduling algorithm, which prioritizes newly arriving requests, enabling more requests to complete within strict deadlines. Nevertheless, \Mname{} still achieves higher SLO satisfaction than NIRVANA.

\begin{table}
\caption{\label{table:ablation-quality} Quality Analysis}
    % \label{table:dataset}
\begin{center}
\scalebox{0.8}{
    \begin{tabular}{ccccc}
        \hline
         Method & \multicolumn{2}{c}{SDXL \cite{podell2023sdxl}} & \multicolumn{2}{c}{SD3 \cite{sd3}} \\
        \hline
          & PSNR($\uparrow$) & SSIM($\uparrow$) & PSNR($\uparrow$) & SSIM($\uparrow$)\\
        \hline
        Distrifusion, 8GPU & 10.96 & 0.49 & 9.35 & 0.38 \\
        \hline
        \makecell[c]{\Mname, \\ Patch Size=128, w/o cache}& {22.13} & {0.77} & {inf} & {1.0} \\
        \makecell[c]{\Mname, \\ Patch Size=256, w/o cache}& {24.84} & {0.81} & {inf} & {1.0} \\
        \makecell[c]{\Mname, \\ Patch Size=512, w/o cache}& {28.82} & {0.88} & {inf} & {1.0} \\
        \hline
        % \makecell[c]{\Mname, \\ Patch Size=256, with cache}& {18.58} & {0.68} & {16.47} & {0.73} \\
        % \hline
    \end{tabular}}
	\end{center}
\end{table}

\begin{figure}[t]
  \begin{center}
    \includegraphics[width=\linewidth]{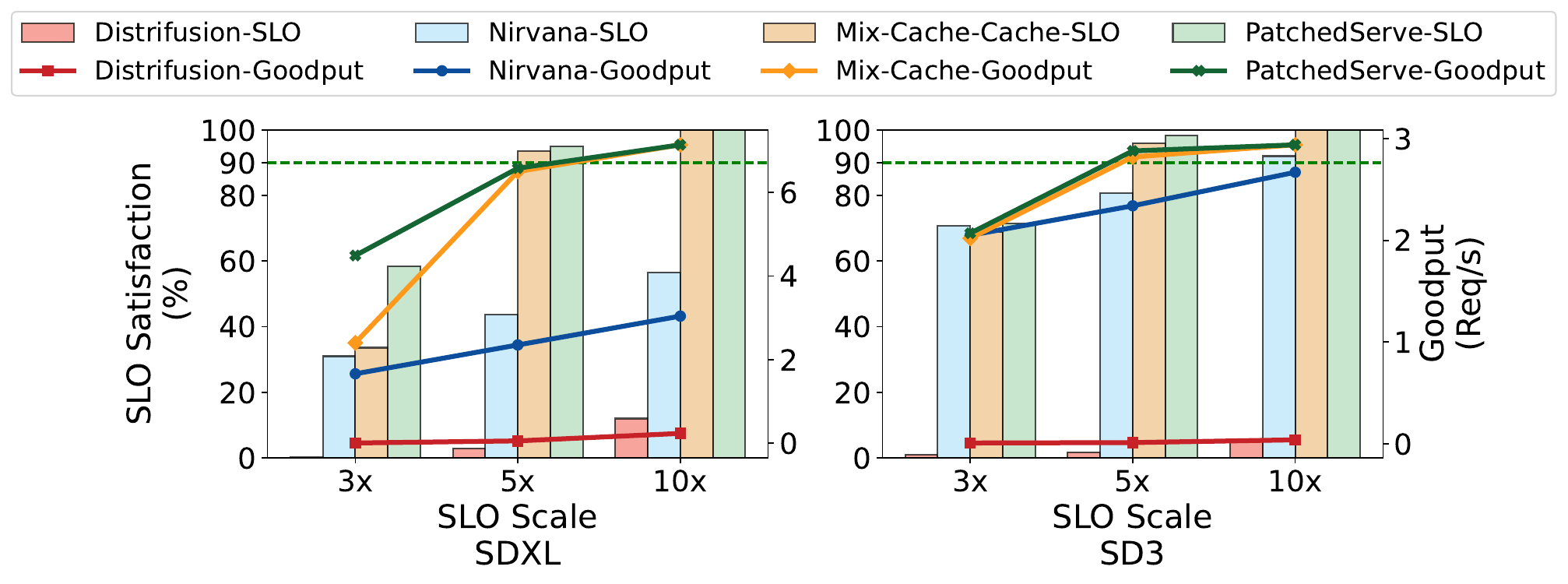}
  \end{center}
  \caption{\label{fig:sensitivity-slo-scale} Performance under various SLO scale.}
\end{figure}

\subsection{Ablation Study}
\label{subsec:analyze}
% \begin{figure}[t]
%   \centering
%     \subfloat[SDXL]{\includegraphics[width=0.5\linewidth]{pic/experiment/sdxl-overhead.jpg}
%     \label{fig:sdxl-overhead}}
%     \subfloat[SD3]{\includegraphics[width=0.5\linewidth]{pic/experiment/sd3-overhead.jpg}
%     \label{fig:sd3-overhead}}
%   \caption{\label{fig:overhead} Latency Overhead from the extra operation.}
% \end{figure}

\begin{figure}[t]
  \begin{center}
    \includegraphics[width=\linewidth]{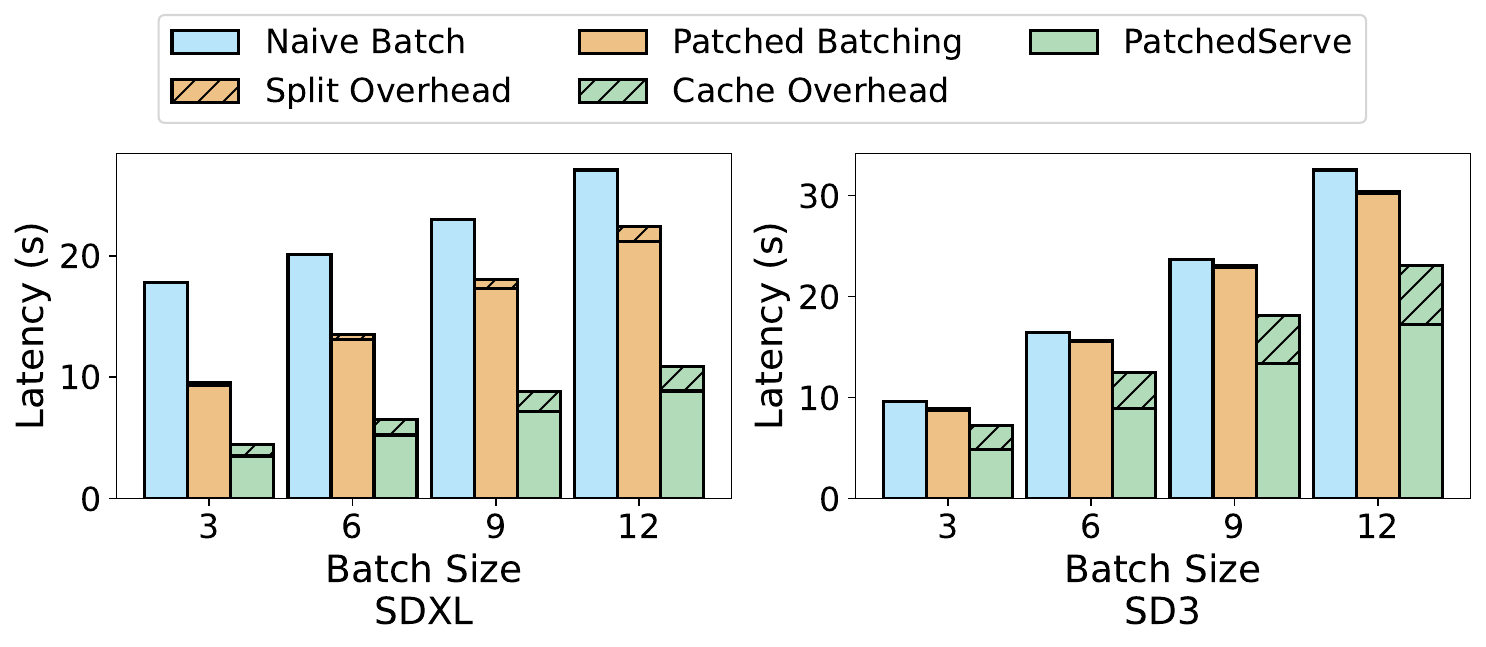}
  \end{center}
  \caption{\label{fig:overhead} Latency Overhead from the extra operation.}
\end{figure}

\textbf{Performance Breakdown}
% Figure \ref{fig:overhead} illustrates the overhead of extra operations brought by \Mname. 
We assess the performance impact of two extra components introduced by \Mname{}, splitting and cache management, by comparing the baseline. We exploit a variant called Patched Batching (which applies only the patch-based batching in \Cref{sec:patch-batch}) and \Mname (which also includes caching from \Cref{sec:patch-cache}) to explore the benefits and overhead.
\Cref{fig:overhead} presents the latency reductions achieved by each technique. A batch size of 3 corresponds to one request per resolution. The baseline benefits from batching only when requests share the same resolution, resulting in rapid throughput gains as batch size increases. Patched Batching depicts an average 13\,\% throughput improvements by processing diverse resolution requests concurrently. The overhead introduced by splitting is minimal, particularly for SD3, which operates on token sequences rather than 2D latent states. SDXL exhibits higher relative improvement than SD3 due to its lower reliance on attention, which limits the benefits of batching. We observe that SD3 incurs higher cache management overhead due to a greater number of blocks per denoising step (24 in SD3 versus 7 in SDXL). Overall, cache management overhead scales modestly with batch size, demonstrating the efficiency of \Mname’s batched cache handling.

\iffalse
\noindent \textbf{Patched Batching Benefits.} We examine the advantages of patched batching to highlight its efficiency. We first show the performance under various patch size settings to demonstrate the reason of our patch size determination, then we compare our patched batch technique with Distrifusion in terms of performance, memory consumption, and quality, to enhance the advantages of our policy.
\fi

% \iffalse
\begin{figure}[t]
  \begin{center}
  \includegraphics[width=\linewidth]{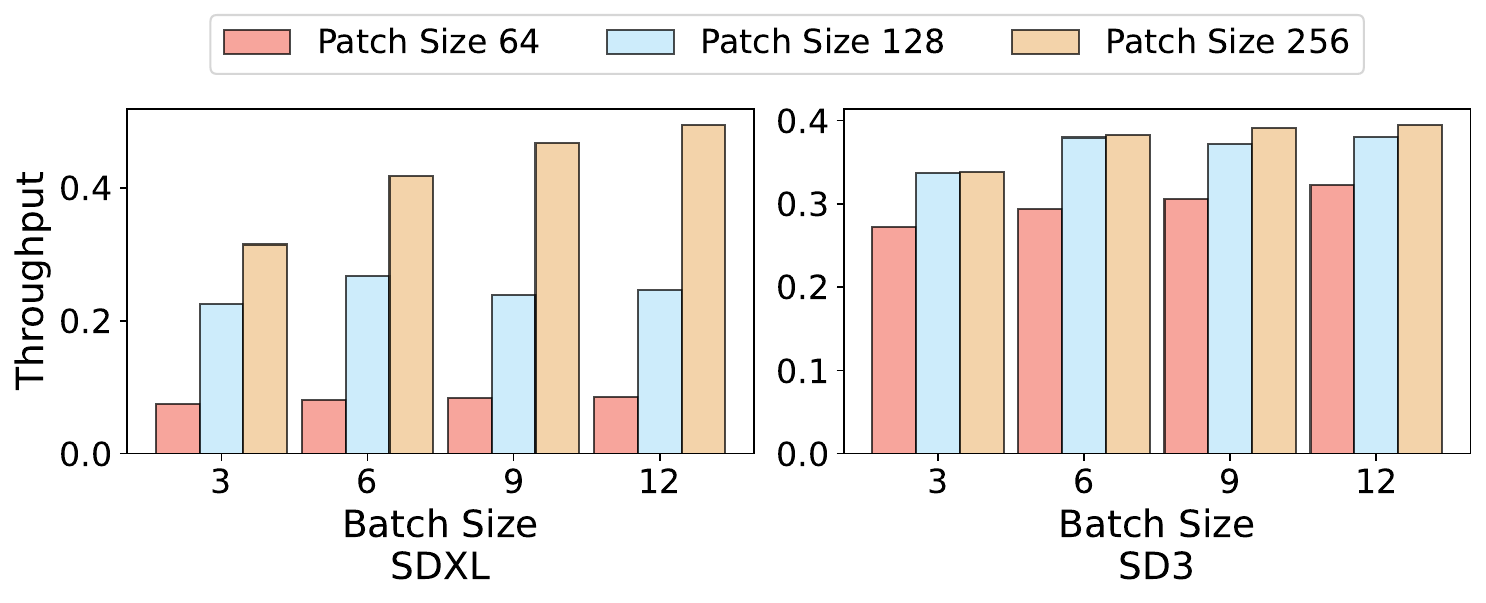}
  \end{center}
  \caption{\label{fig:patch-size-analyze} Average Throughput changing with patch sizes.}
\end{figure}
% \fi

\textbf{Patch size analysis.} We evaluate \Mname{}’s performance across different patch size configurations. As illustrated in \Cref{fig:patch-size-analyze}, throughput increases with the patch size growing, which primarily stems from less splitting overhead, which explains why SDXL manifests a larger decline than SD3 at smaller patch sizes. To mitigate this effect, we configure the patch size as the greatest common divisor of all resolutions within the batch. Additionally, our scheduling algorithm predicts post-batching latency to prevent throughput decline due to unreasonable resolution combination.

\begin{figure}[t]
  \begin{center}
  \includegraphics[width=\linewidth]{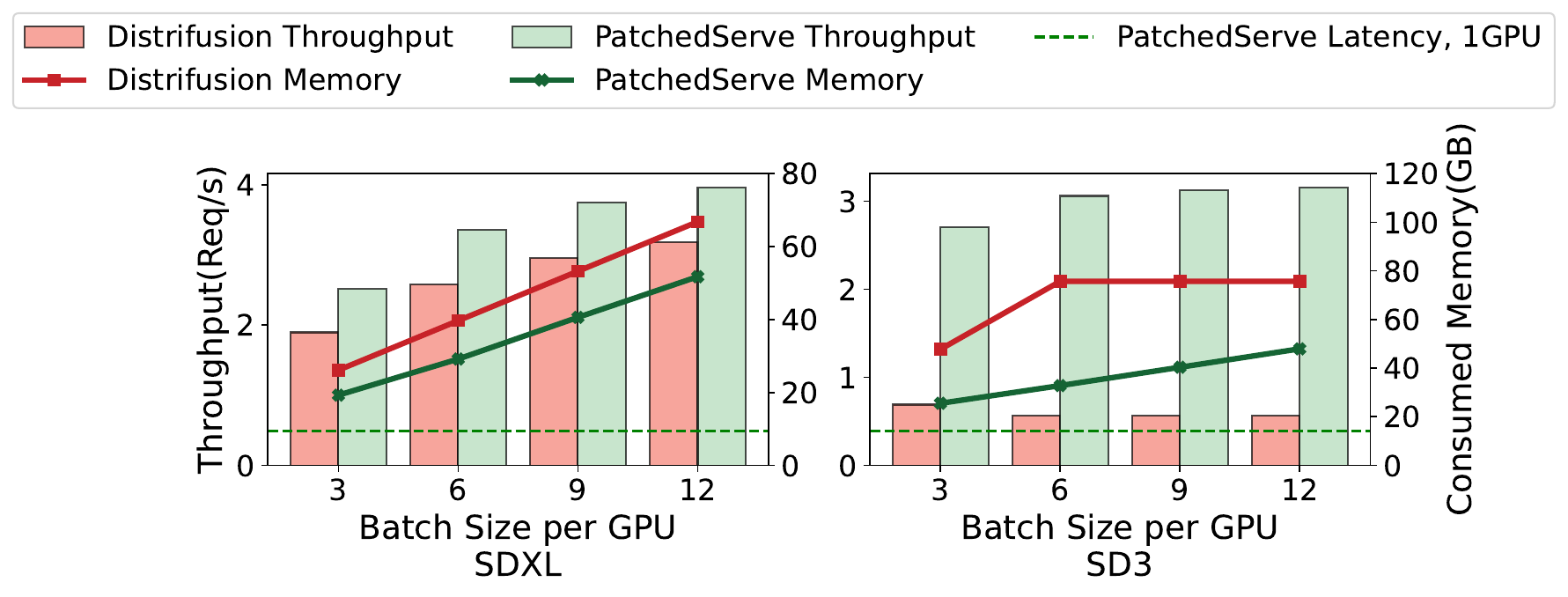}
  \end{center}
  \caption{\label{fig:ablation-distrifusion} Patched batching on Throughput and Memory.}
\end{figure}

\textbf{Comparison with DistriFusion.} \Cref{fig:ablation-distrifusion} presents the average throughput and memory consumption of \Mname{} and DistriFusion. For this evaluation, caching and scheduling are disabled. We evaluate on 8 GPUs, where batch size equals 3 means every GPU has each of three resolution requests on average. \Mname{} dispatches these requests evenly to all 8 GPUs, whereas DistriFusion runs as many requests concurrently as possible. \Cref{fig:ablation-distrifusion} demonstrates that \Mname{}’s throughput across 8 GPUs increases with batch size, reflecting initially low GPU utilization at smaller batch sizes. In contrast, Distrifusion achieves lower throughput on SDXL due to synchronization overhead. Moreover, its communication overhead increases significantly as batch size grows, leading to throughput decreasing on SD3.
\iffalse
In contrast, DistriFusion fails to benefit from larger batch sizes, as it saturates GPU memory at an early stage. This limitation arises because DistriFusion must retain stale KV caches, whereas \Mname{} can operate directly on real-time data. Moreover, its communication overhead increases significantly as batch size grows, leading to throughput decreasing on SD3.
\fi

We further evaluate image quality across varying patch sizes using Peak Signal-to-Noise Ratio (PSNR) \cite{fardo2016formalevaluationpsnrquality} and Structural Similarity Index Measure (SSIM) \cite{ssim} as metrics. PSNR measures differences in pixel intensity, whereas SSIM quantifies similarity between two images. Inf in PSNR indicates 0 pixel-wise difference, and an SSIM of 1.0 denotes  100\,\% structural similarity. We generate 100 $1024\times1024$ images using either DistriFusion or \Mname{}, and compare them against images synthesized by the original model. \Cref{table:ablation-quality} shows that larger patch sizes result in more accurate images generated by \Mname{}. The minor accuracy loss originates from pixel approximations in the Patch Edge Stitcher, whereas the SD3 model achieves 100\,\% accuracy due to the absence of convolution operations. Notably, \Mname{} still attains higher PSNR and SSIM than DistriFusion, owing to its use of up-to-date data rather than stale KV caches.

% \begin{figure}[t]
%   \centering
%     \subfloat[SDXL]{\includegraphics[width=0.5\linewidth]{pic/experiment/sdxl-savings.jpg}
%     \label{fig:sdxl-savings}}
%     \subfloat[SD3]{\includegraphics[width=0.5\linewidth]{pic/experiment/sd3-savings.jpg}
%     \label{fig:sd3-savings}}
%   \caption{\label{fig:savings} Computation savings from patched Vs. full imgs.}
% \end{figure}

\begin{figure}[t]
  \begin{center}
  \includegraphics[width=\linewidth]{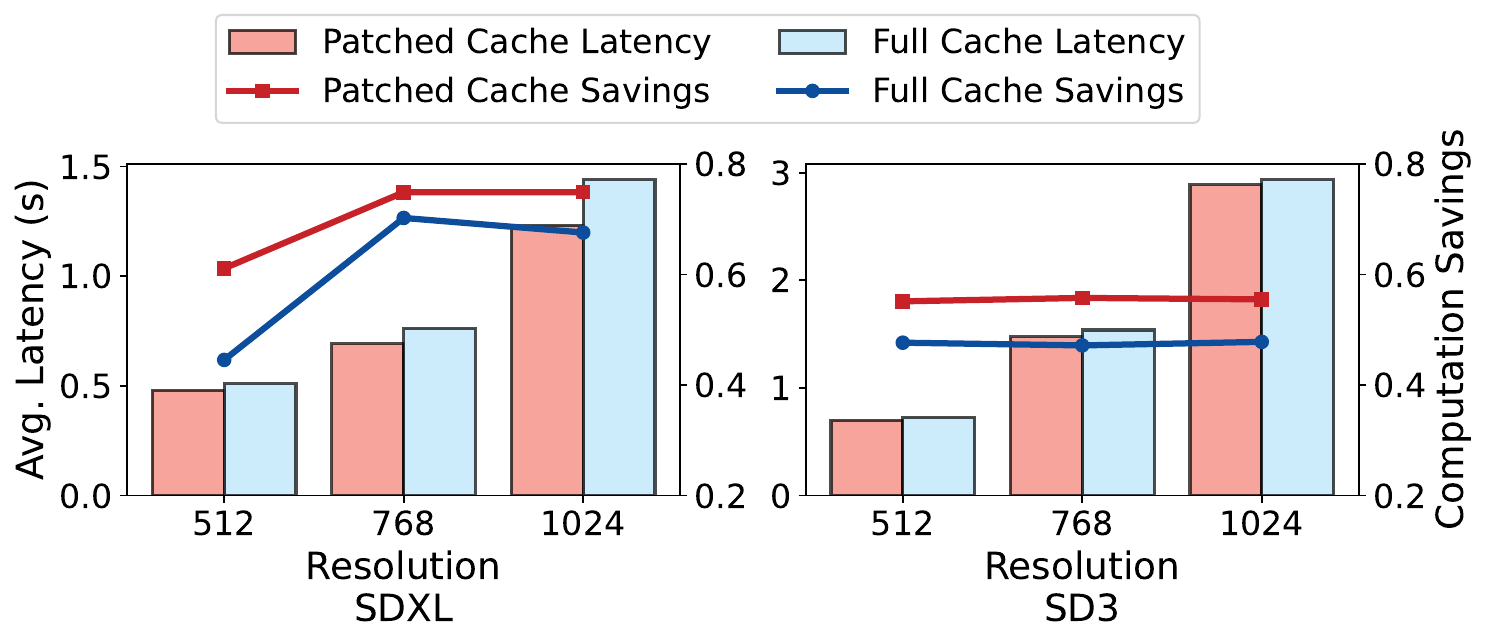}
  \end{center}
  \caption{\label{fig:savings} Computation savings from patched Vs. full imgs.}
\end{figure}

% \textbf{threshold decision}

\noindent \textbf{Caching Benefits}
\label{subsubsec:cache-benefits}
We conduct an experiment to compare the effectiveness of patch-level caching versus whole-image caching. We incorporate whole image caching into Patched Batching Inference and determine that a block can only be skipped if all patches in the current batch meet the similarity threshold. The batch size is set to the maximum capacity supported by our GPU. We measure both the average latency per request and the computational savings, defined as:
${total\_skiped\_patches}$\\/ (${patch\_num \times blocks\_num \times step\_num}$). 
\Cref{fig:savings} demonstrates that patch-level cache reuse consistently outperforms whole-image caching for both models. SD3 exhibits smaller time savings compared to SDXL, reflecting its lower overall computation requirements. Note that the computation savings reported in \Cref{fig:savings} differ slightly from those in \Cref{fig:skip_blocks}, as the latter does not account for cumulative errors. 
\section{Conclusion}
This paper proposes \Mname, the first serving system for hybrid-resolution diffusion models. With the help of patch-based mixed-resolution inference and patch-level cache reuse strategy, \Mname~succeeds in processing requests concurrently regardless of resolutions, achieving better performance. In addition, \Mname~incorporates an SLO-aware schedule algorithm to maximize the number of requests meeting their SLO requirements. In addition, we further prove that our system is easy to scale up to a larger distributed environment and still outperforms the most advanced patch-based diffusion research.

\bibliographystyle{plain}
\bibliography{ref}

\end{document}